\documentclass[aps,pra,showpacs]{revtex4}
\usepackage{amsmath}
\usepackage{amssymb}
\usepackage{bm}
\usepackage{epsfig}
\usepackage{graphicx}
\usepackage{color}
\usepackage{dcolumn}
\usepackage{bm}

\begin{document}
\title{Giant resonant enhancement of optical binding of dielectric particles}
\author{Evgeny N. Bulgakov}
\author{Konstantin N. Pichugin}
\author{Almas F. Sadreev}
\affiliation{Kirensky Institute of Physics, Federal Research
Center KSC SB RAS, 660036 Krasnoyarsk, Russia}
\date{\today}

\begin{abstract}
Optical coupling of two identical dielectric particles gives
rise to bonding and anti-bonding resonances.
 The latter is featured by significant narrowing of the resonant width and strong
enhancement of the $Q$ factor for the high index micron size
particles in subwavelength range.
We consider particles shaped as spheres and disks under coaxial
illumination of dual incoherent counter propagating Bessel beams.
In the case of spheres we derive analytical expressions for the
optical binding (OB) force which decreases as $1/L^2$ for large
distance $L$ between the spheres and displays two periods of
oscillations. For close distances the OB force enormously
increases in the resonant  regime.
 The case of two coaxial disks owing to variation of the distance between disks
  and aspect ratio of each disk is featured by extremal enhancement of the $Q$ factor
compared to the case of two spheres. In that case we demonstrate
unprecedent enhancement of the OB force up to several decades of
nano Newtons. We show that the magnitude and sign of the OB force
strongly depend on the longitudinal wave vector of the Bessel
beams.

\end{abstract}
 \maketitle

\section{Introduction}
The response of a microscopic dielectric object to a light field
can profoundly affect its motion. A classical example of this
influence is an optical trap, which can hold a particle in a
tightly focused light beam \cite{Ashkin1986}. When two or more
particles are present, the multiple scattering between the objects
can, under certain conditions, lead to optically bound states.
This is often referred to peculiar manifestation of optical forces
as optical binding (OB), and   it   was first   observed by Burns
et   al.  on   a   system   of two  plastic   spheres in water  in
1989 \cite{Burns1989}. Optical binding belongs to an interesting
type of mechanical light-matter interaction between particles at
micro-scale mediated by the light scattered by illuminated
particles. Depending on the particle separation, OB leads to
attractive or repulsive forces between the particles and, thus,
contributes to the  formation  of  stable  configurations of
particles. The phenomenon of OB can be realized, for example, in
dual counter propagating beam configurations
\cite{Tatarkova2002,Gomez-Medina2004,Metzger2006,Metzger2006a,Dholakia2010,Bowman2013,Thanopulos2017}.
Equilibrium positions of particles are created by a very weak
balance between the optical forces from the incident fields  and
from the scattered fields generated by the particles. Many
researchers have analyzed OB force quantitatively in theory.
Chaumet {\it et al} \cite{Chaumet2001} and Ng {\it et al}
\cite{Ng2005} calculated the OB force under illumination of two
counter propagating plane waves. \u{C}i\u{z}m\'{a}r et al
\cite{Cizmar2006} presented the first theoretical and experimental
study of dielectric sub-micron particle behavior and their binding
in an optical field generated by interference of two counter
propagating Bessel beams. Also Thanopulos {\it et al}
\cite{Thanopulos2014,Thanopulos2017} numerically evaluated the OB
force as a function of distance between spheres and frequency.

It is clear that excitation of  the resonant modes with high $Q$
factor in dielectric structures results in large enhancement of
near electromagnetic (EM) fields and respectively in extremely
large EM forces proportional to squared EM fields. First, sharp
features in the force spectrum, causing mutual attraction or
repulsion between successive photonic crystal layers of dielectric
spheres under illumination of plane wave has been considered by
Antonoyiannakis and Pendry \cite{Antonoyiannakis1997}. Because of
periodicity of the structure each layer is specified by extremely
narrow resonances which transform into the bonding and anti-boding
resonances for close approaching of the layers. It was shown that
the normal force acting on each layer as well as the total force
acting on both layers including the optical binding force follow
these resonances. It was revealed that the lower frequency bonding
resonance forces push the two layers together and the higher
frequency anti-bonding resonance pull them apart. Later these
disclosures we reported for coupled photonic crystal slabs
\cite{Liu09} and two planar dielectric photonic metamaterials
\cite{Zhang2014} due to existence of resonant states with infinite
$Q$ factor (bound states in the continuum).

However in practice we have arrays of finite number of dielectric
particles which nevertheless show the $Q$ factor exceeding the $Q$
factor of individual particle by many orders in magnitude
\cite{Taghizadeh2017,Sadrieva2019,Bulgakov2019a}. What is
remarkable even two particles can demonstrate  extremely high $Q$
resonant modes owing to avoided crossings. The vivid example is
avoided crossing of whispering-gallery modes (WGM) in coupled
microresonators  which results in extremely high $Q$ factor
\cite{Povinelli2005,Benyoucef2011}. As a result an  enhancement of
the OB force around of hundreds of nano Newtons between coupled
WGM spherical resonators takes place in applied power $1mW$
\cite{Povinelli2005}. However, the WGM modes with extremely high
orbital momenta can be excited only in spheres with large radii of
order $30\mu m$. Recently we offered a solution to the problem of
large $Q$ factor in the subwavelength regime by use of two coaxial
silicon disks of micron sizes. Owing to two-parametric (over the
aspect ratio and distance between disks) avoided crossing of low
order resonances the anti-bonding resonant mode acquires a
morphology of the higher order Mie resonant mode of effective
sphere with extremely small resonant width  \cite{Bulgakov2020}.

In addition to disks we consider  silicon spheres which are
subject to only  one-parameter avoided crossing (the distance
between spheres). As a result the spheres do not show extremely
high $Q$ factors and respectively giant OB forces but have an
advantage of analytical consideration
 of the OB forces  under illumination of dual counter
propagating Bessel beams. We show that two spheres demonstrate the
same features of the OB force which are inherent two disks. The
optical forces for single sphere were explicitly derived by Barton
{\it et al} \cite{Barton1989} in general case that allows to
consider the OB force analytically for the present case of two
spheres. The consideration is significantly simplified when the
spherical particles are subject to beams like Gaussian or Bessel
if they preserve axial symmetry. Then the binding force depends on
the distance between the spheres only
\cite{Ng2005,Metzger2006,Karasek2007,Karasek2009,Zhu2015,Deng2018}.
That allows us to derive analytical expressions for the OB force
which decreases as $1/L^2$ for large distances $L$ between spheres
and displays two periods of oscillations as was first revealed by
Karasek {\it el al} \cite{Karasek2009} numerically. When the
spheres are close to each other the OB force enormously increases
if the frequency of Bessel beams follows to the bonding or
anti-bonding resonances.
 We show also that a
magnitude and what is more interesting the sign of the OB force
strongly depend on the wave number of the Bessel beams that opens
additional options to arrange high index particles optically.
\section{Optical binding force of two spheres}
\begin{figure}
\includegraphics*[width=8cm,clip=]{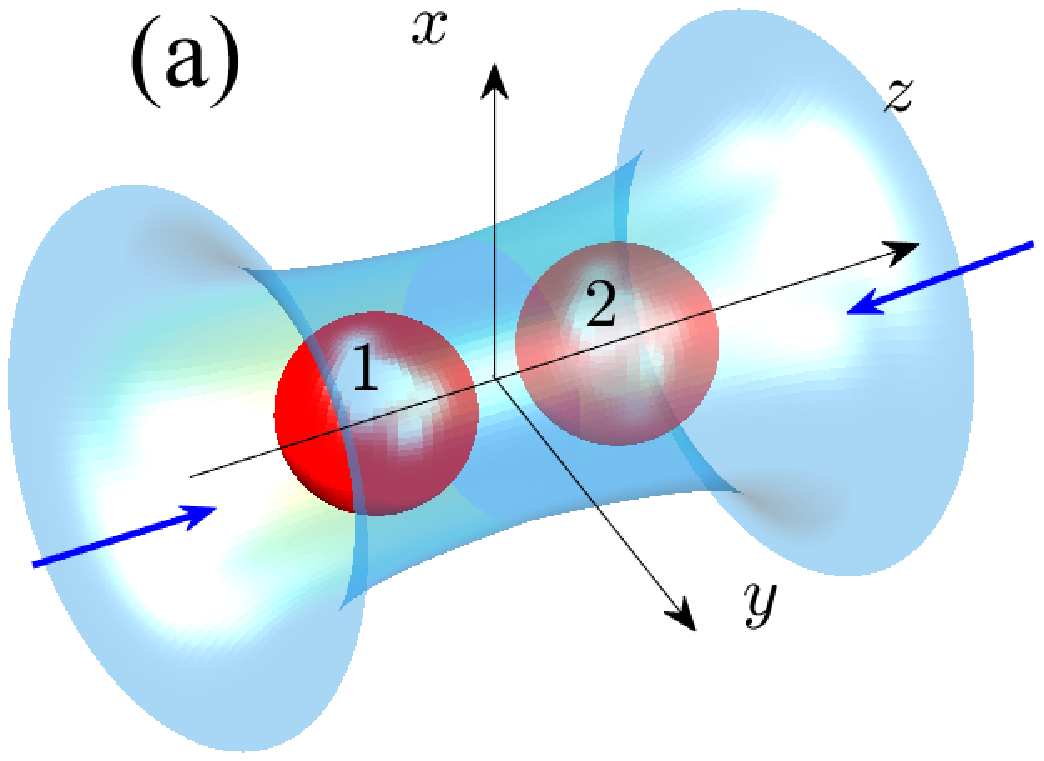}
\includegraphics*[width=8cm,clip=]{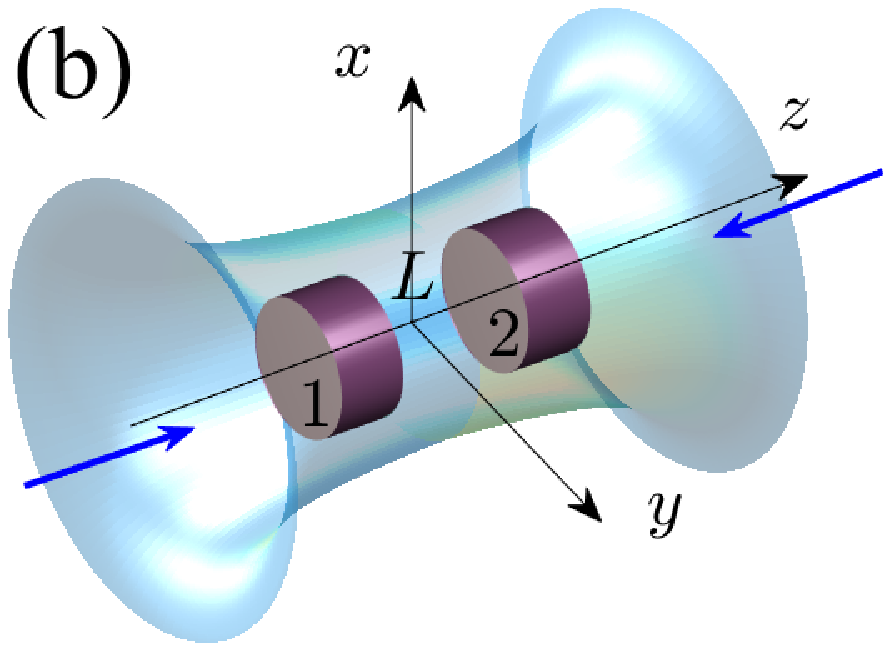}
\caption{ Two silicon spheres (a) and disks (b) with the radius
$a$, permittivity $\epsilon=15$ under illumination of two
counter-propagating mutually incoherent Bessel beams with zero
azimuthal index $m=0$. Light intensity of each beam $P_0=1mW/\mu m^2$.} \label{fig1}
\end{figure}

In order to stabilize the spheres across to beam we use the
results by Milne {\it et al} \cite{Milne2007}  that the Bessel beams strongly trap
spherical particles at the symmetry axis, i.e., at $r=0$ (stable
zero-force points). That justifies the calculation of the OB as
dependent on the distance between the spheres positioned at the
symmetry axis. We consider the Bessel beams with TE polarization
in the simplest form with zero azimuthal index $m=0$
\cite{Karasek2009}
\begin{equation}\label{Bessel}
    {\bf E}_{inc}(r,\phi,z)=E_0{\bf e}_{\phi}\exp(ik_zz)J_1(k_rr)
\end{equation}
where $J_1$ is an Bessel function, $k_z$ and $k_r$ are the
longitudinal and transverse wave numbers, with the frequency
$\omega/c=k=\sqrt{k_r^2+k_z^2}$ and $r, \phi$, and $z$ are the
cylindrical coordinates, ${\bf e}_{\phi}$ is the unit vector of
the polarization. In order to consider the OB force we use the
approach in which two counter-propagating mutually incoherent
Bessel beams were applied \cite{Tatarkova2002,Metzger2006} which
are schematically shown in Fig. \ref{fig1} (a).

The electromagnetic (EM) force is defined by the stress-tensor
$T_{\alpha\beta}$ integrated over the surface elements
$dS_{\beta}$ outside the particle \cite{LL,Antonoyiannakis99}
\begin{eqnarray}\label{force}
&F_{\alpha}=\int T_{\alpha\beta}dS_{\beta}, &\nonumber\\
&T_{\alpha\beta}=
    \frac{1}{4\pi}E_{\alpha}E_{\beta}^{*}-\frac{1}{8\pi}\delta_{\alpha\beta}
|{\bf E}|^2+\frac{1}{4\pi}H_{\alpha}H_{\beta}^{*}-\frac{1}{8\pi}\delta_{\alpha\beta}
   |{\bf H}|^2.& \nonumber
\end{eqnarray}
This problem allows analytical treatment owing to a series of the
Bessel beam and scattered fields both over the vectorial spherical
harmonics. Such an approach was used to find the optical forces
for the case of the isolated sphere
\cite{Chen2009,Wang2013,Song2014,Kiselev2016,Neves2019}. In the
case of two spheres multiple scattering theory was used to define
the OB forces and calculate them numerically
\cite{Ng2005,Metzger2006,Karasek2007,Karasek2009,Zhu2015,Deng2018}.
By using this theory we performed numerical simulations of the
complex resonant frequencies and binding force of two coupled
spheres with focus on the dependence of the OB force on the
intrinsic parameters such as the distance between spheres and
external parameters such the frequency and wave number of the dual
Bessel beams.
\begin{figure}
\includegraphics*[width=10cm,clip=]{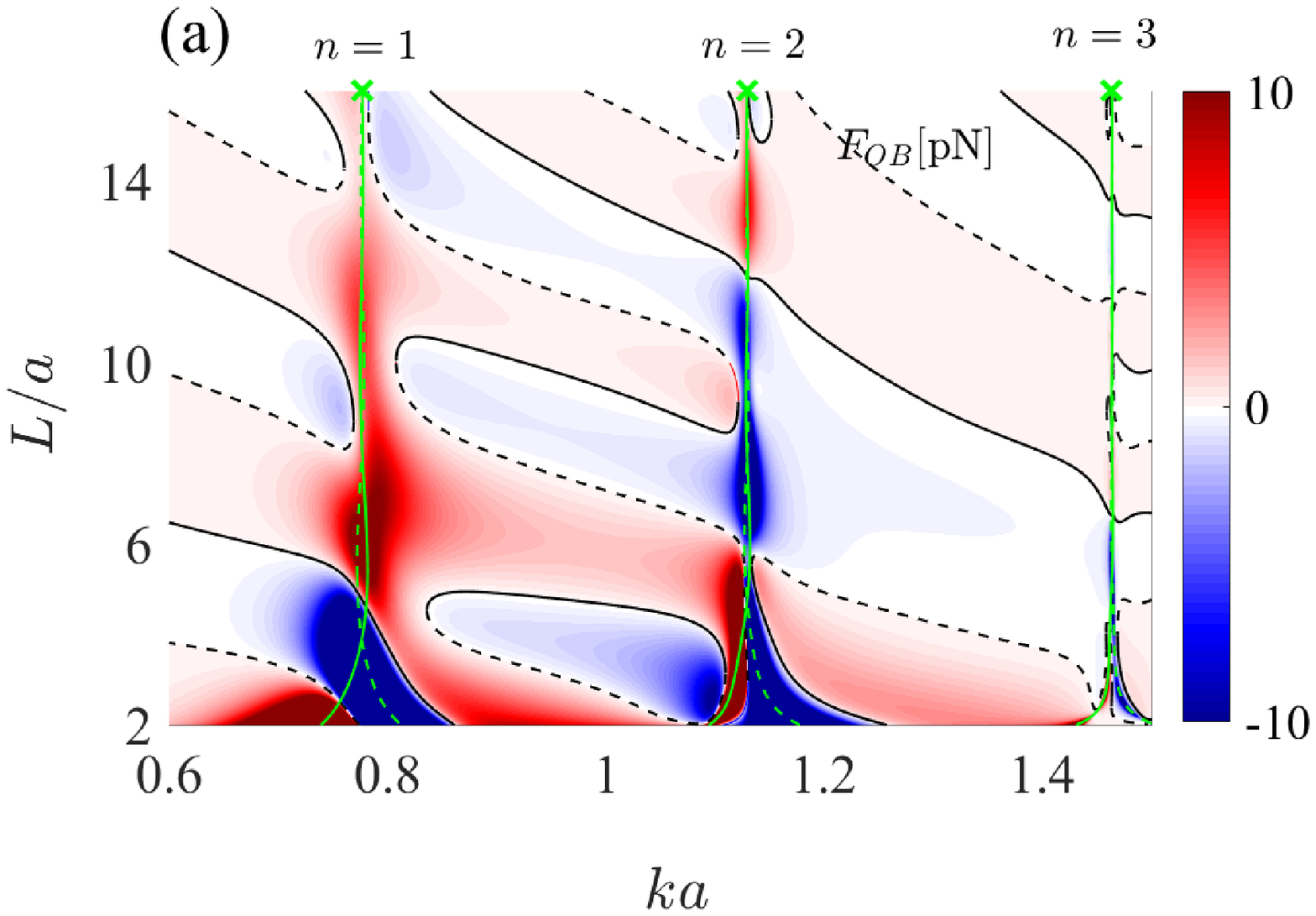}
\includegraphics*[width=8.5cm,clip=]{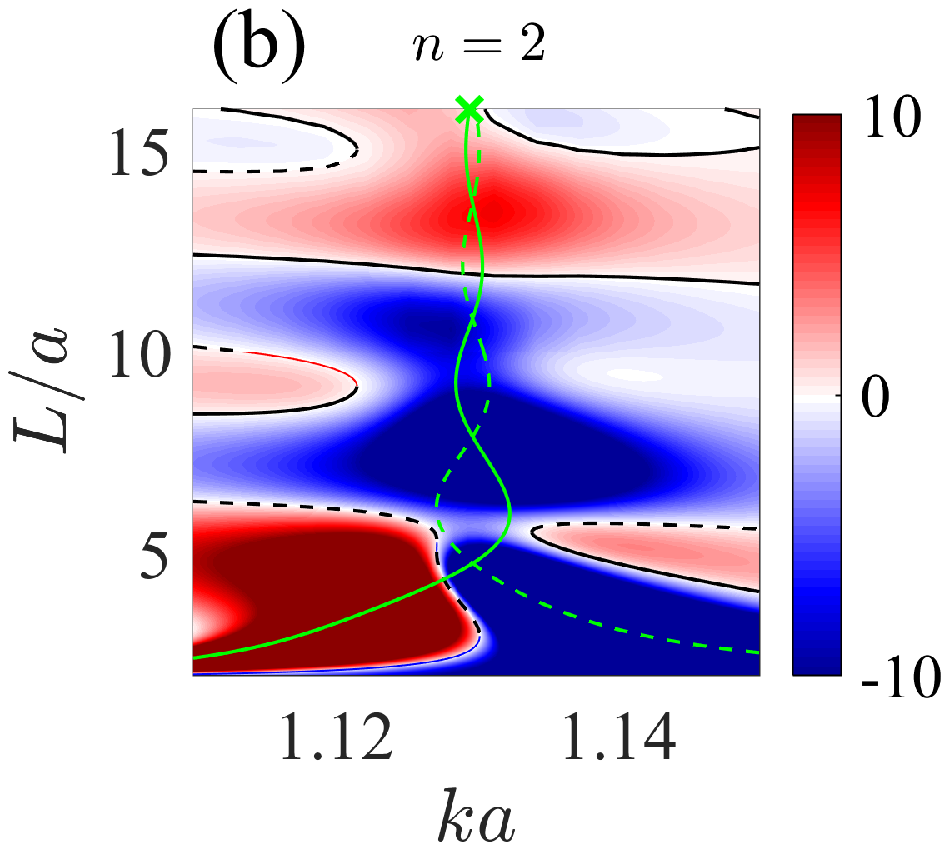}
\includegraphics*[width=8.5cm,clip=]{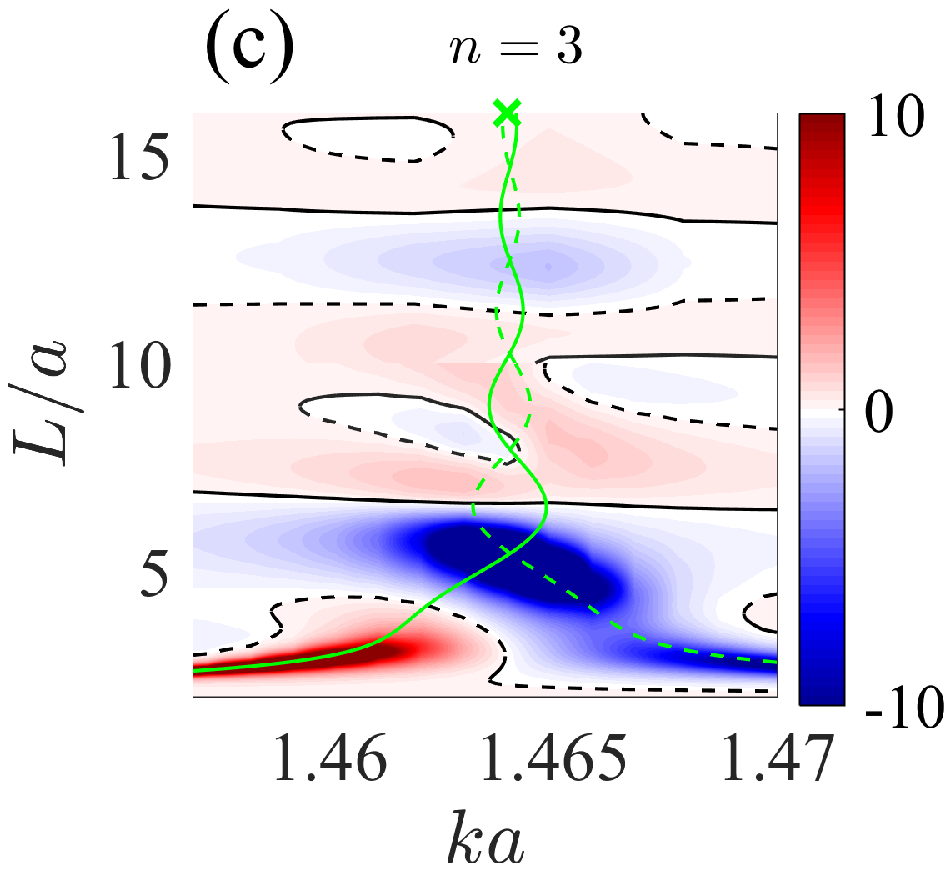}
\caption{The binding force  between two spheres vs the frequency
and distance for the dual Bessel beam of power $1mW/\mu m^2$ with
TE polarization, $k_za=1/2$ where $a=0.5\mu m$ is the sphere
radius with $\epsilon=15$. (b) and (c) zoomed versions. The red
corresponds to attractive forces and blue corresponds to repulsive
OB force. Black solid (dash) lines show stable (unstable)
configuration of spheres. Light green solid (dash) lines show
symmetric (anti symmetric) resonant frequencies of two spheres vs
the distance between. Crosses mark the Mie TE resonances in
isolated dielectric sphere.} \label{fig2}
\end{figure}
\begin{figure}
\includegraphics*[width=10cm,clip=]{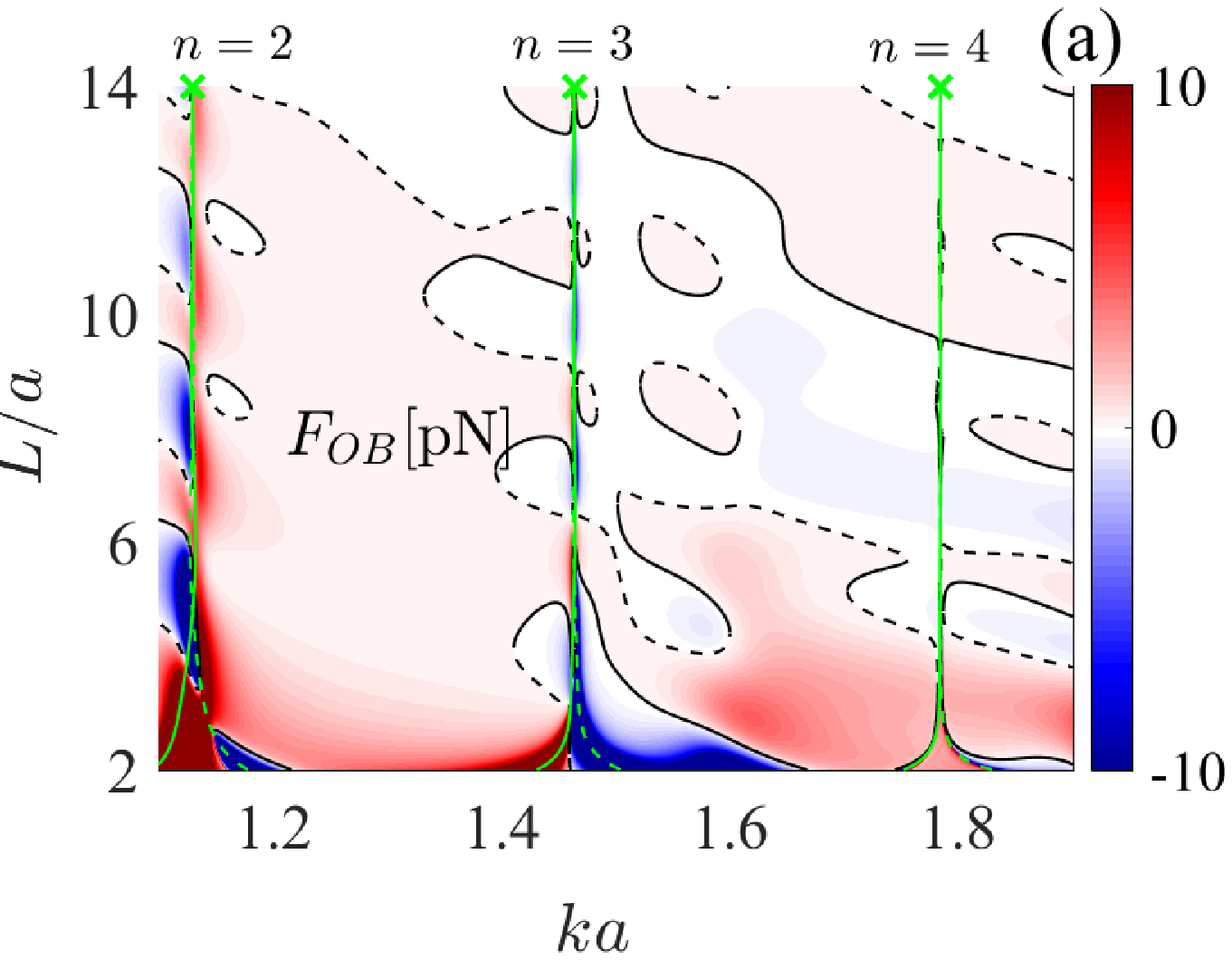}
\includegraphics*[width=8.5cm,clip=]{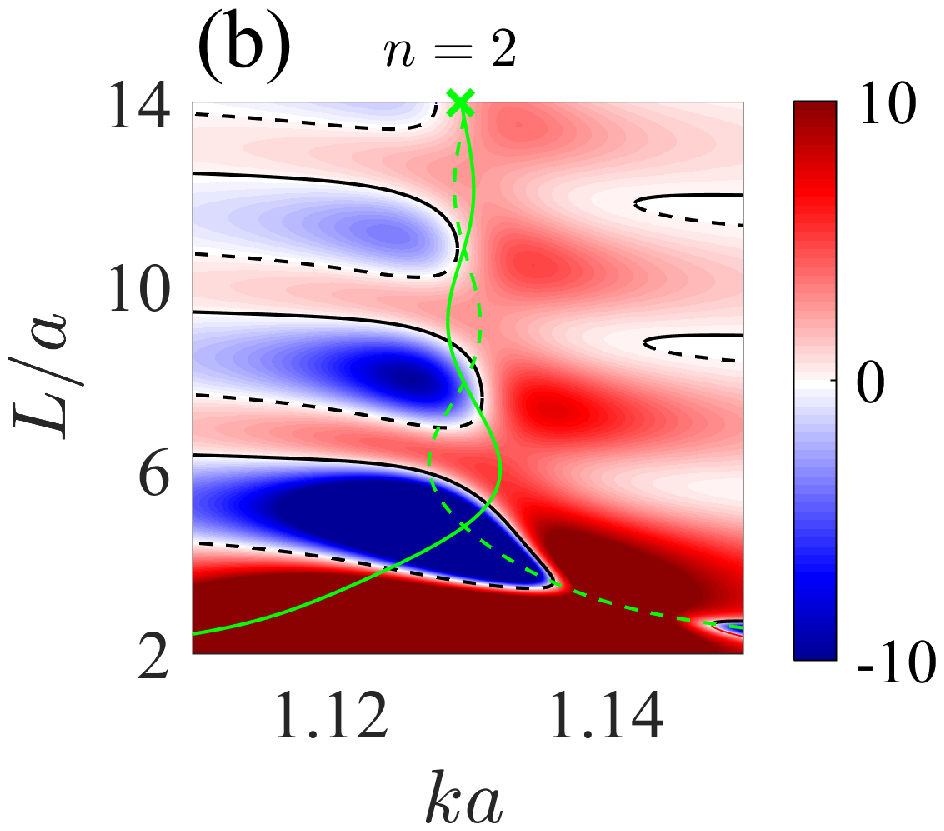}
\includegraphics*[width=8.5cm,clip=]{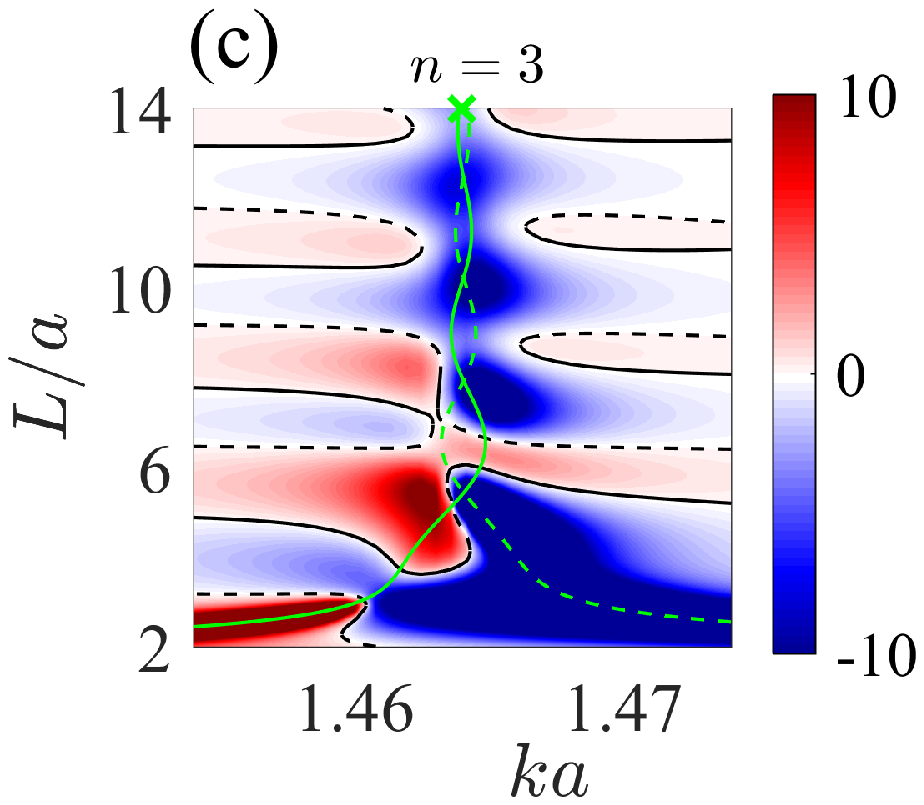}
\caption{The same as in Fig. \ref{fig2} but for $k_za=1$.}
\label{fig3}
\end{figure}

The results of calculations are presented in Figs. \ref{fig2} and
\ref{fig3} for two values $k_za=1/2$ and $k_za=1$ of the Bessel
beam (\ref{Bessel}). We show the binding force
$F_{OB}^{\rightarrow}=(F_{1z}-F_{2z})/2$ where the indices 1 and 2
denote the spheres where the Bessel beam incident at the left.
Owing to an incoherence of the Bessel beam illuminated from the
right we have the same expression for
$F_{OB}^{\rightarrow}=-F_{OB}^{\leftarrow}$. As a result we obtain
doubled value for the OB force $F_{OB}=F_{1z}-F_{2z}$. Strong
resonant forces above $10 pN$ by absolute value in Figs.
\ref{fig2} and \ref{fig3} are saturated by intense red
(attractive) or blue (repulsive).  As it was expected the OB force
shows vivid resonant behavior near the Mie resonances of the
individual sphere labelled by orbital index $n$ because of
resonant enhancement of scattered fields.
However, for variation of the distance between spheres we see a
number of peculiarities. The first one is periodic oscillations
when the repulsive OB force is alternating by the attractive one.
Respectively, the equilibrium distances shown in Figs. \ref{fig2}
and \ref{fig3} by solid lines follow a law $(k+k_z)L =2\pi
l+\phi_0, l=1, 2, 3, \ldots $ and undergo abrupt changes near the
Mie resonances $k_n$. This period will be derived below
asymptotically for large $L$ and first was predicted by Karasek
{\it el al} \cite{Karasek2009}. Also in Figs. \ref{fig2} and
\ref{fig3} we show the resonant frequencies of two spheres
(even/odd relative to $z\rightarrow -z$ or bonding/antibonding)
versus the distance between spheres which will be analyzed below
by the use of multiple scattering theory. The second peculiarity
is the decrease of the resonant OB force with the order of the Mie
resonance $n$. We consider roughly that the more $n$ is the less
is the coupling of the Bessel beam for given $k_z$ with the Mie
resonant mode.

As it was said above the case of two spheres enables analytical
treatment of the OB force in the resonant approximation. Owing to
the axial symmetry of the total system of two spheres and applied
Bessel beam we can take $m=0$ with only three components of EM
field $E_{\phi}, ~H_r, ~H_{\theta}$ for TE polarization in
spherical system. Then outside the spheres the EM fields scattered
by the spheres can  be presented as a series in the vectorial
spherical harmonics as follows \cite{Stratton}
\begin{equation}\label{M1M3}
{\bf E}({\bf r})=\sum_n\sum_{j=1,2}b_{n}^{(j)}{\bf M}_{n0}^{(3)}({\bf r-r_j})
\end{equation}
where ${\bf r}_j$ are positions of centers of spheres,
\begin{equation}\label{jh}
{\bf M}_{n0}^{(3)}({\bf
r})={\bf X}_{n0}(\theta,\phi)h_n(kr),
\end{equation}
where ${\bf X}_{n0}(\theta,\phi)$ are the vector spherical
harmonics \cite{Linton2013},  $h_n(z)$ are the Hankel functions.
Here and below the angular index $m=0$ in $b_{n0}^{(j)}$ is
omitted. For the case of the single sphere the optical forces were
explicitly derived by Barton {\it et al} \cite{Barton1989} in
general case. For the present particular case $m=0$ the z-th
component of optical force acting on the first sphere equals
\begin{equation}\label{F}
    F_{z}^{(1)} = -F_0{\rm Im}\sum_nf_n
    [2b_{n+1}^{(1)}b_n^{(1)*}+b_{n+1}^{(1)}B_n^{(1)*}+b_n^{(1)*}B_{n+1}^{(1)}]
\end{equation}
where $b_n^{(1)}$ are the coefficients of series (\ref{M1M3}),
$F_0=\frac{a^4k^2E_0^2}{4\pi},~~
f_n=\frac{n(n+1)(n+2)}{\sqrt{(2n+1)(2n+3)}}$ and
\begin{equation}\label{bn1}
b_n^{(j)}=S_n(k)B_n^{(j)}, j=1,2,
\end{equation}
where the so-called Lorenz-Mie coefficients are given by
\begin{equation}\label{S_l}
S_n(k)=\frac{j_n(\sqrt{\epsilon}ka)[rj_n(kr)]_{r=a}'-j_n(ka)[rj_n(\sqrt{\epsilon}kr)]_{r=a}'}
{h_n(ka)[rj_n(\sqrt{\epsilon}kr)]_{r=a}'-j_n(\sqrt{\epsilon}ka)[rh_n(kr)]_{r=a}'},
\end{equation}
$j_n(x)$ are the spherical Bessel functions. The case of two
spheres was developed by Thanopulos {\it et al}
\cite{Thanopulos2014}. In contrast to Ref. \cite{Thanopulos2014}
we reveal that the OB force is basically focused around the Mie
resonances for high index particles. That prompts to use the
resonant approximation which substantially simplifies analysis
because of elimination of sum over the orbital indices $n$. Thus,
we can truncate the series in Eq. (\ref{F}) with a preservation of
only resonant term given by the Lorenz-Mie coefficient $S_n(k)$
around the n-th Mie resonance. We have for the optical force
acting on the $j$-th sphere
\begin{equation}\label{Fn1}
    \frac{F_{zn}^{(j)}}{F_0}\approx \sum_{\sigma=\pm 1}(-1)^{(1+\sigma)/2}f_{n+(\sigma-1)/2}{\rm Im}
    [S_n(k)^{*}B_{n+\sigma}^{(j)}B_n^{(j)*}],~~\sigma=\pm 1.
\end{equation}

The incident fields radiating the first sphere
is superposed of the incident Bessel beam and the field scattered by the second sphere
\begin{eqnarray}\label{Bn}
&B_n^{(1)}=B_n^{(inc)}+B_{n,n}^{(21)}(L)b_n^{(2)},&\nonumber\\
&B_{n+\sigma}^{(1)}=B_{n+\sigma}^{(inc)}+B_{n,n+\sigma}^{(21)}(L)b_n^{(2)}.&
\end{eqnarray}
Due to the addition theorem \cite{Mackowski1991,Linton2013} we can write the contribution of the second sphere
as follows
\begin{eqnarray}\label{AT}
    &B_{n,n}^{(21)}(L)=4\pi\sum_{p=0,2, 4, \ldots}^{2n}
    g_{nnp}i^p\mathcal{G}(n0;n0;p)Y_p^0(1)h_p(kL),&\nonumber\\
    &B_{n,n+1}^{(21)}(L)=4\pi\sqrt{\frac{n}{(n+2)}}\sum_{p=1,3, 5,
    \ldots}^{2n+\sigma}g_{n,n+1,p}i^p\mathcal{G}(n0;n+1,0;p)\hat{}Y_p^0(1)h_p(kL),&
\end{eqnarray}
where the argument $1$ of Legandre polynomials is related to the
direction of scattered field.
Substituting the specific expressions for the spherical Bessel functions
 into Eq. (\ref{AT}) with account of coefficients $g$ and $\mathcal{G}$
 tabulated in Ref. \cite{Linton2013} we find for $kL\gg 1$
\begin{equation}\label{A_ll}
B_{n,n}^{(21)}(L)\approx -c_{n,n}\frac{e^{ikL}}{(kL)^2}, ~~
B_{n,n+\sigma}^{(21)}(L)\approx -ic_{n,n+\sigma}\frac{e^{ikL}}{(kL)^2},
\end{equation}
where $c_{1,1}=3, c_{2,2}=15, c_{1,2}=6.708,  c_{2,3}=25.1, \ldots$ are
real coefficients.

For the large $kL$ we can rewrite Eq. (\ref{Bn}) as follows
\begin{eqnarray}\label{Bn1}
&B_n^{(1)}\approx B_n^{(inc)}(1+B_{n,n}^{(21)}(L)S_n(k)e^{ik_zL}),&\nonumber\\
&B_{n+\sigma}^{(1)}\approx
B_{n+\sigma}^{(inc)}+B_{n,n+\sigma}^{(21)}(L)S_n(k)B_n^{(inc)}e^{ik_zL}.&
\end{eqnarray}
where we took into account that the Bessel beam (\ref{Bessel})
accumulates the phase factor $e^{ik_zL}$ when reaches the second
sphere. Substituting here asymptotes (\ref{A_ll}) and using an
inequality $|B_n^{(inc)}|\gg |B_{n,n}^{(21)}(L)b_n^{(2)}|$ we can
approximate
\begin{eqnarray}\label{Bnapprox}
&B_n^{(1)}\approx B_n^{(inc)}[1-\frac{S_n(k)}{(kL)^2}c_{n,n}e^{i(k_z+k)L}],&\nonumber\\
&B_{n+\sigma}^{(1)}\approx
B_{n+\sigma}^{(inc)}-\frac{iS_n(k)}{(kL)^2}c_{n,n+\sigma}B_n^{(inc)}e^{i(k_z+k)L},&
\end{eqnarray}
As a result we obtain the following expression for the optical force (\ref{Fn1}) onto the first sphere
\begin{eqnarray}
\label{FF1n}
&    F_{zn}^{(1)}\approx F_0\sum_{\sigma=\pm 1}(-1)^{(1+\sigma)/2}f_{n+(\sigma-1)/2}
[{\rm Im}(S_n^{*}B_n^{(inc)*}B_{n+\sigma}^{(inc)})&,\nonumber\\
-&\frac{c_{n,n}}{(kL)^2}|S_n^{2}B_n^{(inc)}B_{n+\sigma}^{(inc)}|\sin((k+k_z)L+\phi_{n+\sigma})-
\frac{c_{n,n+\sigma}}{(kL)^2}|B_n^{(inc)}S_n|^2\cos(k+k_z)L].&
\end{eqnarray}
where
$\phi_{n+\sigma}=Arg(B_n^{(i)}B_{n+\sigma}^{(inc)*}S_n(k)^{2})$.
Similarly, we have for the second sphere
\begin{eqnarray}\label{Bn2}
&B_n^{(2)}=B_n^{(inc)}e^{ik_zL}+B_{n,n}^{(12)}(L)b_n^{(1)},&\nonumber\\
&B_{n+\sigma}^{(2)}=B_{n+\sigma}^{(inc)}e^{ik_zL}-B_{n,n+\sigma}^{(12)}(L)b_n^{(1)}.&
\end{eqnarray}
By use of  identities
\begin{equation}\label{ident}
B_{n,n}^{(21)}=B_{n,n}^{(12)}, ~B_{n,n+\sigma}^{(21)}=-B_{n,n+\sigma}^{(12)}
\end{equation}
and Eq. (\ref{A_ll})
we can rewrite Eq. (\ref{Bn2}) as follows
\begin{eqnarray}\label{Bn22}
&B_n^{(2)}=B_n^{(inc)}[e^{ik_zL}-\frac{c_{n,n}}{(kL)^2}S_ne^{ikL}],&\nonumber\\
&B_{n+\sigma}^{(2)}=B_{n+\sigma}^{(inc)}e^{ik_zL}+\frac{ic_{n,n+\sigma}}{(kL)^2}S_nB_n^{(inc)}e^{ikL}.&
\end{eqnarray}
As a result we have for the force acting on the second sphere
\begin{eqnarray}
\label{FF2n}
&    F_{zn}^{(2)}\approx F_0\sum_{\sigma=\pm 1}(-1)^{(1+\sigma)/2}f_{n+(\sigma-1)/2}
[{\rm Im}(S_n^{*}B_n^{(inc)*}B_{n+\sigma}^{(inc)})&,\nonumber\\
+&\frac{c_{n,n}}{(kL)^2}|S_n^{2}B_n^{(inc)}B_{n+\sigma}^{(inc)}|\sin((k-k_z)L-\phi_{n+\sigma})+
\frac{c_{n,n+\sigma}}{(kL)^2}|B_n^{(inc)}S_n|^2\cos(k-k_z)L],&
\end{eqnarray}
i.e.,
\begin{equation}\label{symkz}
F_{zn}^{(2)}(k_z)=-F_{zn}^{(1)}(-k_z).
\end{equation}
Therefore, the asymptotes at $kL\gg 1$ for OB force  owing to the
dual Bessel beams propagating along the z-axis equal
\begin{eqnarray}
\label{OB} & F_{OB}(L)=F_{zn}^{(1)}(k_z)-F_{zn}^{(2)}(k_z)\approx
 F_0\sum_{\sigma=\pm 1}(-1)^{(1+\sigma)/2}f_{n+(\sigma-1)/2}&\nonumber\\
&\frac{c_{n,n}}{(kL)^2}|S_n^{2}B_n^{(inc)}B_{n+\sigma}^{(inc)}|[\sin((k+k_z)L+\phi_{n+\sigma})
+\sin((k-k_z)L+\phi_{n+\sigma})]+&\nonumber\\
&\frac{c_{n,n+\sigma}}{(kL)^2}|B_n^{(inc)}S_n|^2[\cos(k+k_z)L+\cos(k+k_z)L].&
\end{eqnarray}

This expression shows two properties of the OB for long distances
between spheres: the long-distance and short-range modulation of
the binding force $\frac{2\pi}{k-k_z}$ and $\frac{2\pi}{k+k_z}$
that was reported by Karasek {\it el al} \cite{Karasek2009}
numerically by use of a coupled dipole method. It is worthy to
note that first the oscillatory behavior of the OB was observed
already by Burns {\it et al} \cite{Burns1989} that was used for
separation of $1.43\mu m$ polystyren particles in water. An
asymptotical decline $1/L^2$ of the OB force can be also
understood if we consider the scattered field from the second
sphere positioned  at the the z-axis at the distance $L$ is given
by the vector spherical function \cite{Linton2013}
\begin{equation}\label{M}
{\bf M}_{n0}({\bf r}-{\bf e}_zL)=-{\bf e}_{\phi}h_n(kL)
\frac{dP_n^0(\cos\theta)}{d\theta}.
\end{equation}
For integration over the first sphere positioned at the $z=0$ the
contribution of the second sphere is proportional to
$\sin\theta=a/L$. As a result together with the asymptotic of the
Bessel function $h_l(kL)\sim\frac{e^{ikL}}{kL}$ we obtain the
total asymptotic $1/L^2$. We notice that this asymptotic is
justified only for coaxial illumination of spheres by the Bessel
beams.

\begin{figure}
\includegraphics*[width=10cm,clip=]{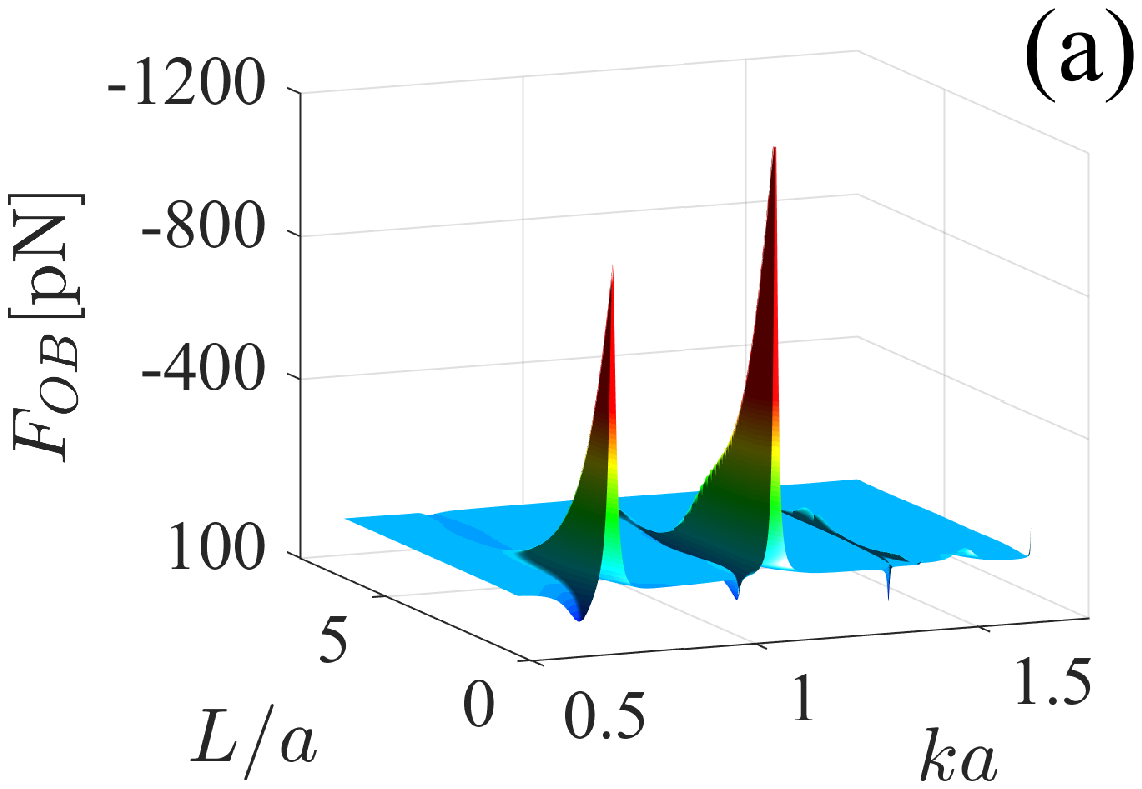}
\includegraphics*[width=10cm,clip=]{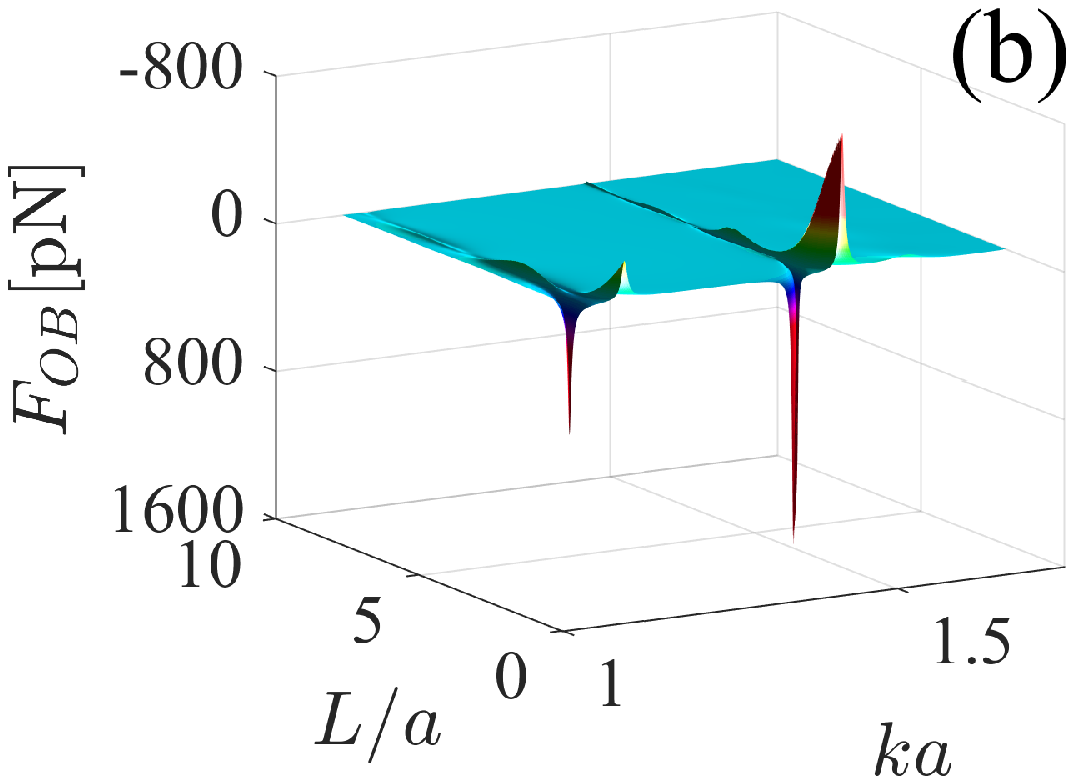}
\caption{The binding force between two spheres vs the wavelength
and distance for (a) $k_za=1/2$ and (b) $k_za=1$.} \label{fig4}
\end{figure}

The behavior of OB at the close vicinity of spheres $L\rightarrow
2a$ is more dramatic as Fig. \ref{fig4} demonstrates.
In order to analytically evaluate this behavior we employ the
multiple scattering theory which reduces the Maxwell equations
into the linear algebraic equations for the amplitudes $b_n$ in
expansions of EM fields over vectorial spherical harmonics
(\ref{M1M3}) given by the index $n$ and $m=0$ which can be written
as matrix equation
\begin{equation}\label{scat}
    \widehat{L}(k)\vec{\psi}=\vec{\psi}_{inc}
\end{equation}
where the non Hermitian, non symmetric  matrix $\widehat{L}(k)$ is
determined by a specific structure of dielectric particles. The
incident state $\vec{\psi}_{inc}$ is the column of amplitudes
$B_n^{(inc)}$ in this representation. The resonances are given by
the solutions of the homogeneous equation
\begin{equation}\label{BSC}
    \widehat{L}(k)\vec{\psi}=0
\end{equation}
for complex eigenvalues $k$ whose real parts are shown by solid
and dash lines in Figs. \ref{fig2} (b) and (c). For the present
problem it is important to note that the matrix $\widehat{L}(k)$
can be defined in the basis of left and right eigenvectors
\begin{equation}\label{bio}
\vec{y}_{\lambda}\widehat{L}(k)=\lambda\vec{y}_{\lambda}, ~~
\widehat{L}(k)\vec{x}_{\lambda}=\lambda\vec{x}_{\lambda},
\end{equation}
where $\vec{y}_{\lambda}\vec{x}_{\lambda'}=\delta_{\lambda,\lambda'}$.
By use the condition of completeness
\begin{equation}\label{compl}
    \sum_{\lambda}\vec{x}_{\lambda}\vec{y}_{\lambda}=1
\end{equation}
we write the following equalities
\begin{equation}\label{AA}
\widehat{L}(k)=\sum_{\lambda}\lambda\vec{x}_{\lambda}\vec{y}_{\lambda},~~
\widehat{L}^{-1}(k)=\sum_{\lambda}\frac{\vec{x}_{\lambda}\vec{y}_{\lambda}}{\lambda}.
\end{equation}
as well as for the solution of Eq. (\ref{scat})
\begin{equation}\label{aa}
  \vec{\psi}=\sum_{\lambda}\frac{W_{\lambda}}{\lambda}\vec{x}_{\lambda}
\end{equation}
where
\begin{equation}\label{coupl}
    W_{\lambda}=\vec{y}_{\lambda}\vec{\psi}_{inc}
\end{equation}
are the coupling coefficients of the incident wave with the
eigenmodes of the open system. For the case of high refractive
index of dielectric sphere the index $\lambda$ can be related to
those resonant term which has the smallest $\lambda_n$ in the
vicinity of the resonant frequency $k\approx {\rm Re}(k_n)$.
That allows to  write in the vicinity of the $n$-th Mie resonant
frequency ${\rm Re}(k_n)$ the eigenvalue as
\begin{equation}\label{lambda}
    \lambda_n=q_n(k-k_n).
\end{equation}

\begin{figure}
\includegraphics*[width=7cm,clip=]{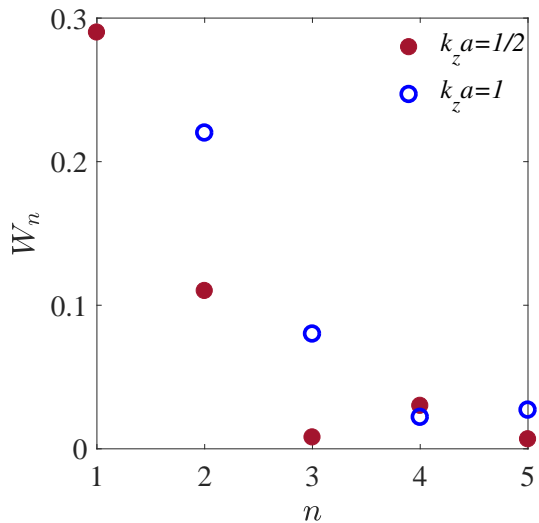}
\caption{The coupling coefficient $W_n(k,k_z)$ vs the order of Mie
resonance $n$.} \label{Wn}
\end{figure}

For the case of identical high index particles resonant modes can be presented as symmetric (bonding) and
anti-symmetric (anti-bonding) modes \cite{Pichugin2019}
\begin{eqnarray}\label{hybridl0}
&{\bf E}_{n}\approx \frac{E_0W_{n,s}}{(k-k_{n,s})}{\bf E}_{n,s}+
\frac{E_0W_{n,a}}{(k-k_{n,a})}{\bf E}_{n,a},&
\end{eqnarray}
where the factors $1/q_n$ are absorbed by the coupling constants
$W_{n,s,a}$,
\begin{equation}\label{hybrid}
{\bf E}_{n,s,a}({\bf r})\approx \frac{1}{\sqrt{2}}[{\bf M}_{n0}({\bf r}-\frac{L}{2}
{\bf e}_z)\pm {\bf M}_{n0}({\bf r}+\frac{L}{2}{\bf e}_z)]
\end{equation}
$E_0$ is the amplitude of the Bessel beam. The coupling constant
of incident Bessel beam (\ref{Bessel}) with the symmetric or
anti-symmetric resonant modes (\ref{hybrid}) can according to the
definition (\ref{coupl}) be presented as (see also
\cite{Song2019})
\begin{equation}\label{W}
    W_{n,s,a}\approx
  W_{n}(k,k_z)\left\{\begin{array}{cc}\cos k_zL/2\\
   i\sin k_zL/2   \end{array}\right.
\end{equation}
where $W_{n}(k,k_z)$ is the coupling constant of the Bessel beam
with the $n$-th Mie resonant mode. One can perform analytical
calculations of the constant by the use of a great deal of algebra
presented in Refs. \cite{Nes2007,Jiang2012,Kiselev2016,Neves2019}.
However it is simpler to find the coupling constants numerically
because their values are independent on the distance. The results
are presented in Fig. \ref{Wn} for $k_za=0.5, 1$ and show
that the OB force decreases with $n$.

\begin{figure}
\includegraphics*[width=5cm,clip=]{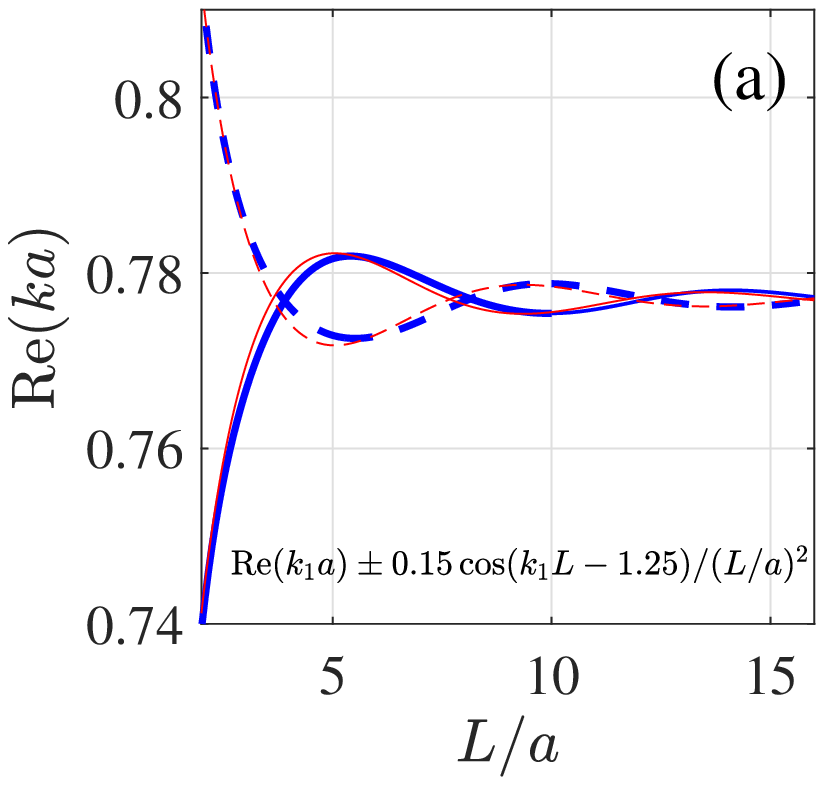}
\includegraphics*[width=5cm,clip=]{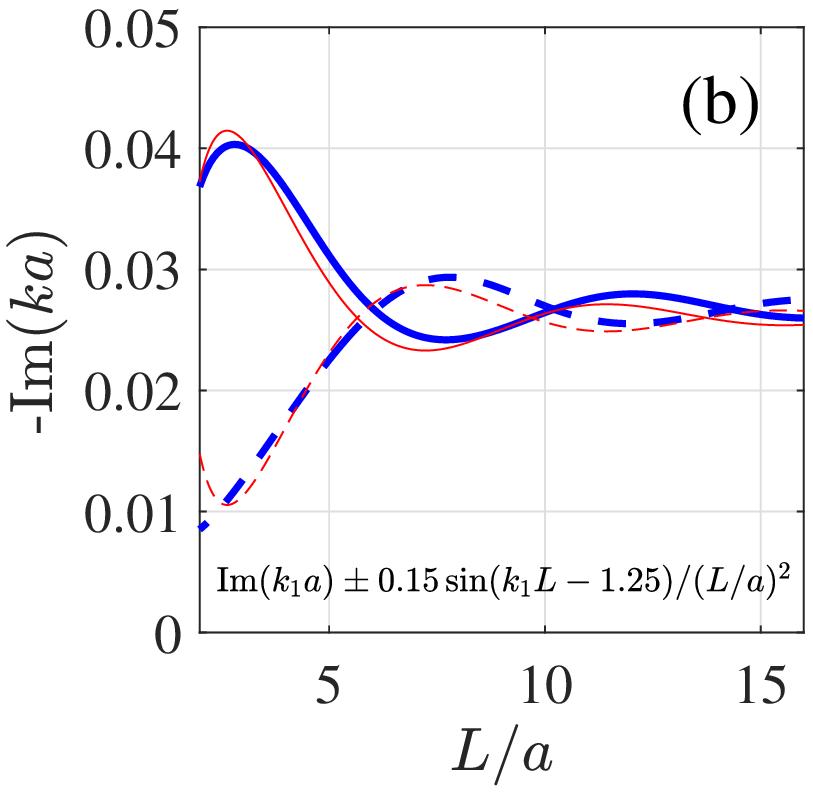}
\caption{The dipole $n=1$  resonant frequencies of two coupled
spheres, real (a) and imaginary (b) vs the distance calculated
from Eq. (\ref{BSC}) (thick blue) and compared to two-level
approximation (\ref{k_sa}) (thin red) and presented with fitting
parameters in Figures. Solid (dash) lines show the frequencies of
bonding (anti-bonding) resonances.} \label{fig5}
\end{figure}

The resonant frequencies in the two-level approximation can be
written as follows \cite{Pichugin2019}
\begin{equation}\label{k_sa}
    k_{n,s,a}={\rm Re}(k_{n,s,a})-i\gamma_{n,s,a}
    \approx k_n\pm \frac{v_n}{L^2}e^{i(k_nL-\theta_n)}.
    \end{equation}
Fig. \ref{fig5} shows that the resonant frequencies (\ref{k_sa})
well describes numerically calculated
dipole resonances $n=1$  with fitting parameters $v_1=0.15$ and
$\theta_1=1.25$. As seen from Figs. \ref{fig2} and \ref{fig3} at
close distances between spheres the bonding and anti-bonding
resonances are well separated that allows us to consider them
independently. In what follows we consider  in details the
anti-bonding dipole resonance $n=1$ for which the OB noticeably
exceeds the case of the bonding resonance as Fig. \ref{fig4} (a)
shows. The reason is related to the denominators in Eq.
(\ref{hybridl0}) which equal the imaginary parts of the resonances
${\rm Im}(k_{n,s,a})$ at $k={\rm Re}(k_{n,s,a})$. Other words, the
near fields are proportional to
 the quality factors $Q_{n,s,a}=-{\rm Re}(k_{n,s,a})/2{\rm Im}(k_{n,s,a})$.
 For the dipole case resonances with $n=1$
the $Q_{1,s}\rightarrow 10$ while $Q_{1,a}\rightarrow 56$ at
$L\rightarrow 2a$. The response of the scattered field around the
anti-bonding resonance becomes strong compared to the incident
Bessel beam.  Therefore the incident field can be neglected. Fig.
{\ref{fig6} (a) demonstrates that the scattered field indeed
slightly differs from the antisymmetric mode $\vec{E}_{n,a}$ given
by Eq. (\ref{hybrid}). That directly correlates with the behavior
of the resonant width vs $L$ shown in Fig. \ref{fig5} (b). One can
see that ${\rm Im}(k_{1,a})$ has a minimum at $L\approx 2a$. While
Fig. \ref{fig6} (b) shows that the Bessel beam contributes
significantly into the scattered field when  $k\approx k_{1,s}$
and therefore can not be disregarded. That is a consequence of the
resonant width of the anti-bonding dipole resonant mode $1,s$. One
can see from Fig. \ref{fig5} (b) that the resonant width of the
bonding dipole resonant mode $1,s$ reaches maximum for
$L\rightarrow 2a$.
\begin{figure}
\includegraphics*[width=12cm,clip=]{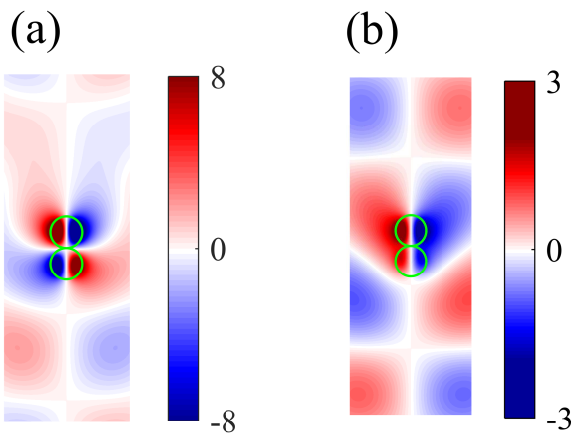}
\caption{Numerically computed scattered field (the component
$E_{\phi}$) at the closest distance $L=2a$ for frequencies around
the dipole Mie resonance $k_1$: (a) $k={\rm Re}(k_{1,a})$
(anti-bonding resonant mode),  (b) $k={\rm Re}(k_{1,s})$ (bonding
resonant mode)
The Bessel beam with $k_za=0.5$ illuminates spheres from bottom.}
\label{fig6}
\end{figure}

Comparison of Eq. (\ref{hybridl0}) with Eq. (\ref{bn1}) gives us
\begin{equation}\label{b1b2}
    b_n^{(j)}=(-1)^{j-1} E_0d_n, j=1,2,
\end{equation}
where $d_n=\frac{W_{n,a}}{\sqrt{2}q\gamma_{n,a}}\sin (k_zL/2)$.
Owing to Eqs. (\ref{bn1}) and (\ref{Bn}) we have
\begin{eqnarray}\label{BBn1}
&B_n^{(1)}\approx d_n(\frac{1}{S_n(k_n)}-B_{n,n}^{(21)}(L)), &\\
&B_{n+1}^{(1)}\approx B_{n,n+1}^{(21)}(L)b_n^{(2)}=-d_nB_{n,n+1}^{(21)}(L)&.\nonumber
\end{eqnarray}
According to Eq. (\ref{F}) we obtain for the force acting onto the first sphere around the anti-bonding
dipole resonance $k\approx Re(k_{1,a})$:
\begin{equation}\label{F11}
    \frac{F_{z1}^{(1)}}{F_0}\approx -{\rm Im}(S_1^{*}B_1^{(1)*}B_2^{(1)})\approx
    -|d_1|^2{\rm Im}[(1-S_1B_{1,2}^{(21)*}(L))(1+B_{1,1}^{(21)*}(L))]
\end{equation}
where
$$B_{1,1}^{(21)}(L)=h_0(kL)+h_2(kL), ~B_{1,2}^{(21)}(L)=-1.3416(h_1(kL)+h_3(kL))$$
 owing to Eq. (\ref{AT}).
Taking into account relations (\ref{ident}) we obtain the OB at $L\geq 2a$
\begin{equation}\label{OBL2}
    \frac{F_{OB}}{F_0}=2\frac{F_{z1}^{(1)}}{F_0}=
    \frac{1.3416f_1|W_1|^2\sin^2(k_zL/2)}{\gamma_{1,a}^2(L)}
    {\rm Im}[(1-S_1^{*}(k_{1,a}))(h_0^{*}(kL)+h_2^{*}(kL))(h_1(kL)+h_3(kL))].
\end{equation}

Fig. \ref{fig7}(a) shows the asymptotic formula (\ref{OBL2})
perfectly describes the numerically computed OB force for the
dipole anti-bonding resonance.
\begin{figure}
\includegraphics*[width=7cm,clip=]{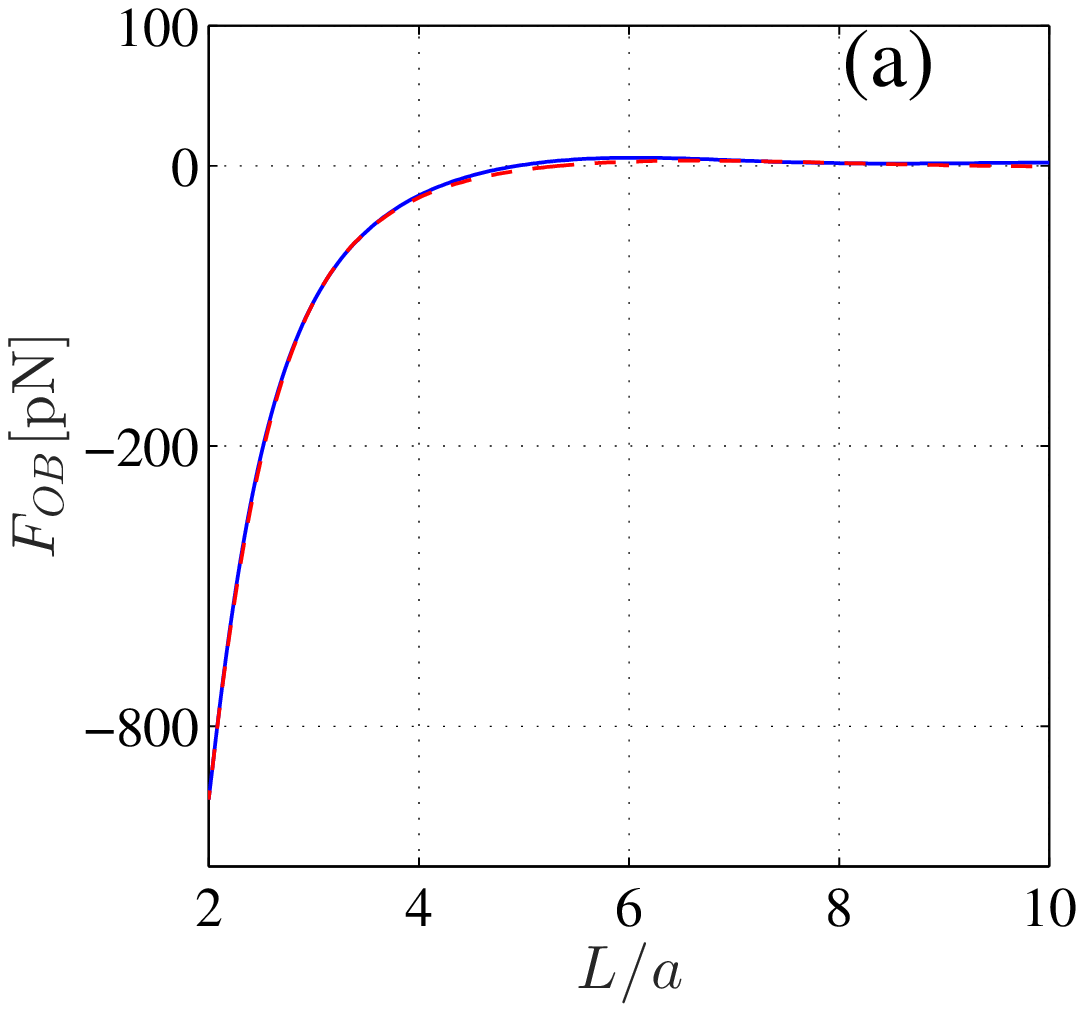}
\includegraphics*[width=7cm,clip=]{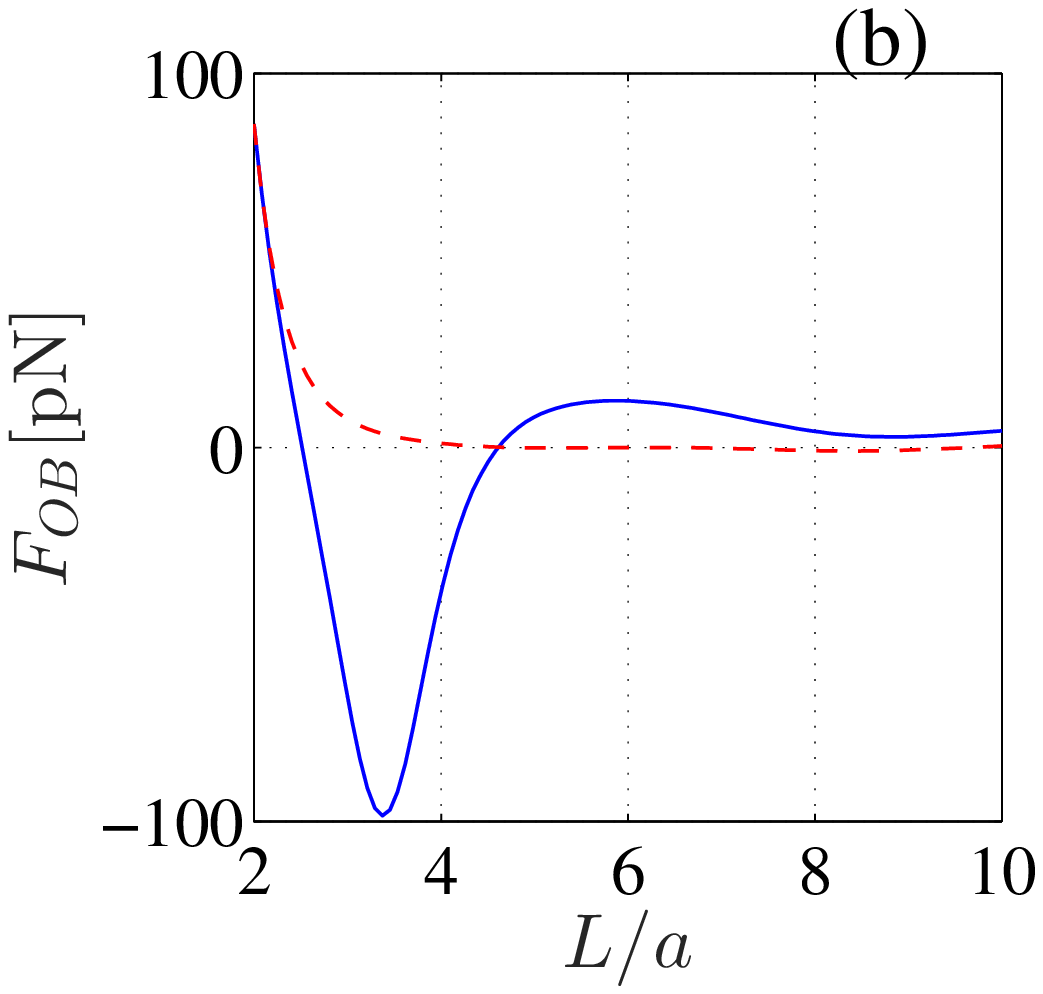}
\caption{The OB vs distance between spheres at the vicinity of the
dipole anti-bonding (a) resonance $k={\rm Re}(k_{1,a}(L))$ and (b) bonding
resonance $k={\rm Re}(k_{1,s}(L))$. The solid line shows numerics and dash
line shows approximated formulas.  Wave number of Bessel beam
$k_za=1/2$.} \label{fig7}
\end{figure}
Similar asymptotic formula can be written for the bonding
resonance by simple substitution $a\rightarrow s$. However  Fig.
\ref{fig7} (b) shows strong discrepancy between numerics and
asymptotic formula. The discrepancy is related to that as seen
from Fig. \ref{fig5} (b) the bonding resonant width reaches
maximum at $L\rightarrow 2a$. As a result enhancement of scattered
EM field at the bonding resonance roughly four times yields to the
case of anti-bonding resonance. Therefore for calculation of the
optical forces we can not neglect the incident fields as distinct
from the dipole anti-bonding resonance.

Next,  with growth of the order of the Mie TE resonances $n$ in
the dielectric sphere the resonant width exponentially decreases
\cite{Lam1992,Gorodetsky1996}. Therefore one could expect the fast
growth of the OB force. However by the same reason of reduction of
radiation losses with $n$ decrease of the coupling of the Mie
resonant modes with the Bessel beam occurs that Fig. \ref{Wn}
demonstrates.

Moreover two parameters, the frequency and by wave vector $k_z$
along the propagation axis $z$  define the Bessel beam
(\ref{Bessel}). Figs. \ref{fig2} and \ref{fig3} show that indeed
these parameters noticeably effect the equilibrium distances
between the spheres. Eq. (\ref{OBL2}) predicts simple dependence
of the OB on the longitudinal wave number $k_z$ of the Bessel beam
in the form of $\sin^2k_zL/2$ but rather complicated dependence on
the distance $L$ through the Hankel functions for the anti-bonding
dipole resonance. This conclusion is illustrated in Figs.
\ref{fig8} which shows strong dependence of the OB force on $k_z$
and $L$ for frequencies tuned to the dipole and quadruple
anti-bonding frequencies ${\rm Re}(k_{1,a})$ and ${\rm
Re}(k_{2,a})$, respectively. One can see that these results
provide potentially useful way to manipulate distance between
particles by variation of the longitudinal wave number of the
Bessel beams.
\begin{figure}
\includegraphics*[width=8cm,clip=]{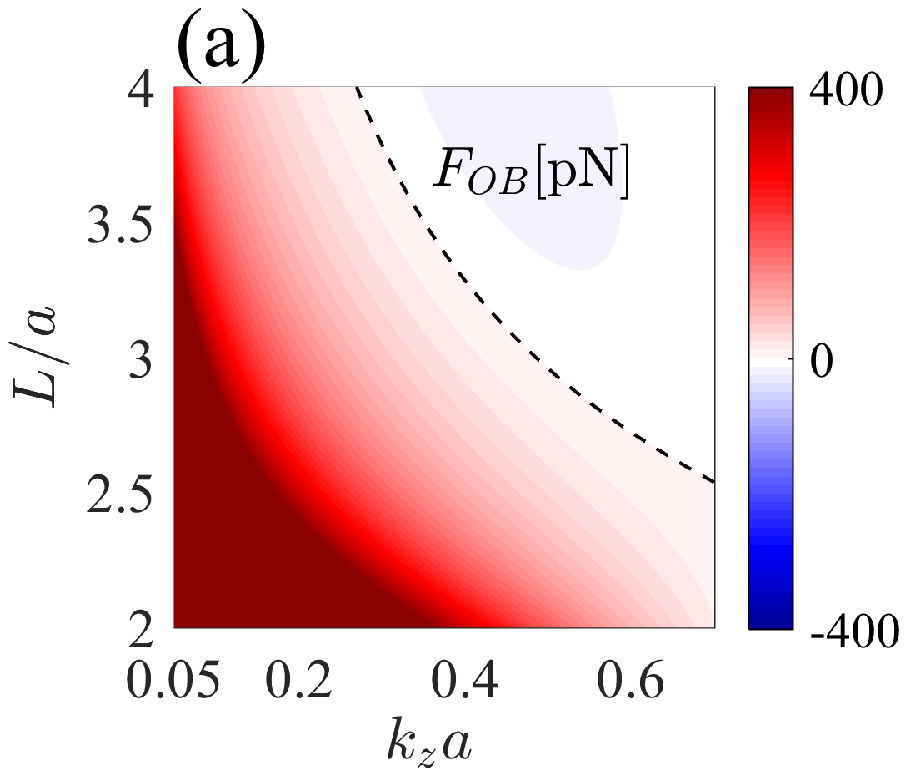}
\includegraphics*[width=8cm,clip=]{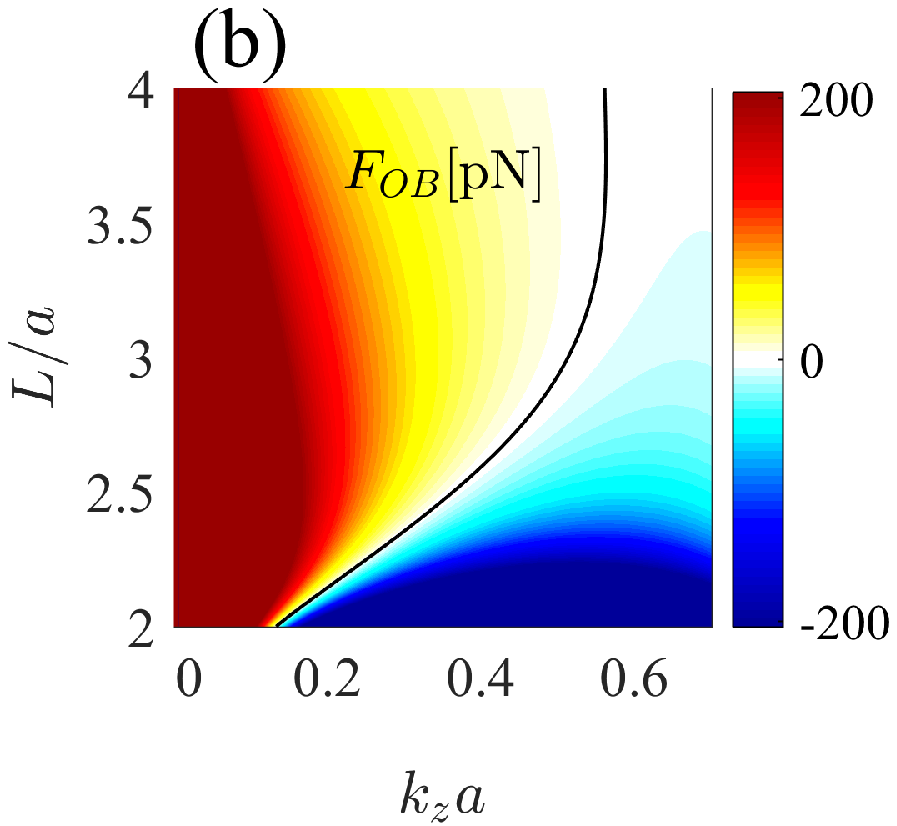}
\includegraphics*[width=8cm,clip=]{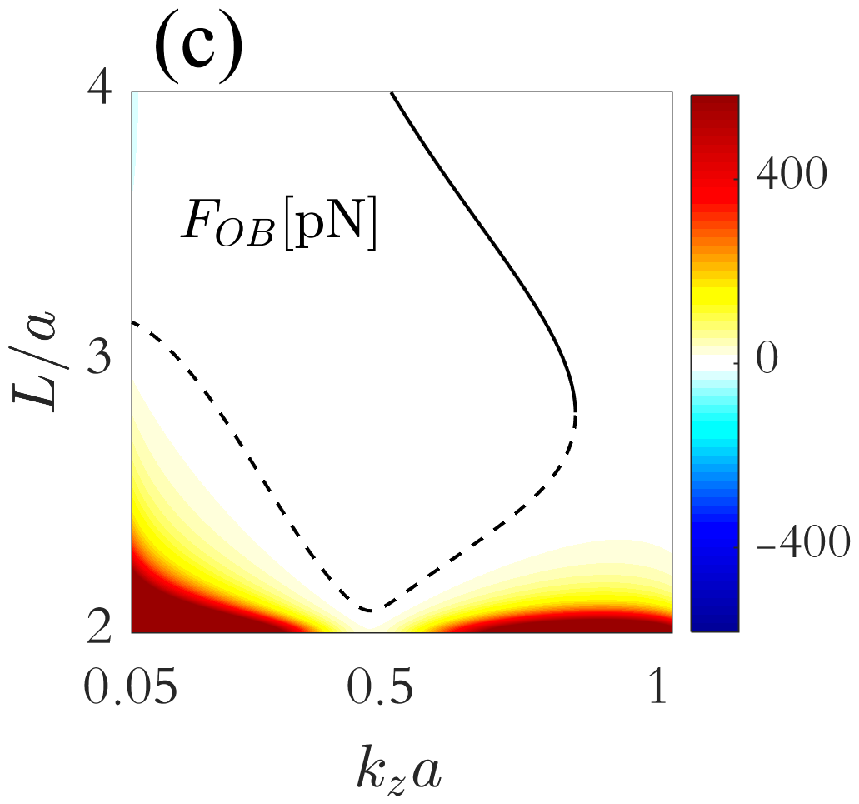}
\includegraphics*[width=8cm,clip=]{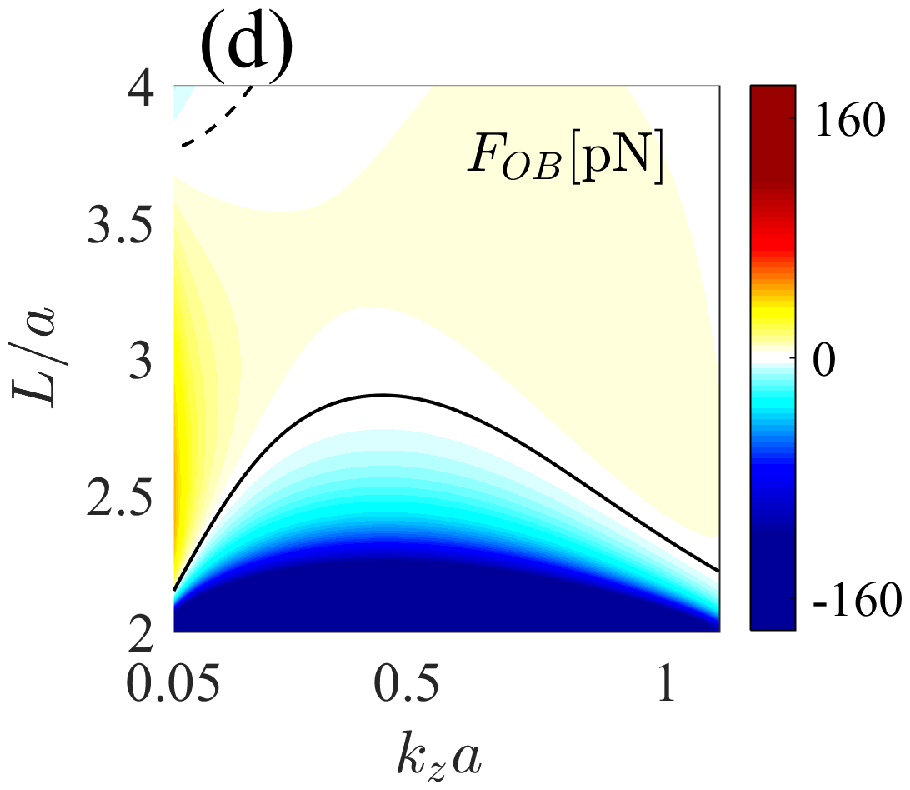}
\caption{The OB vs distance between spheres and longitudinal wave
number of the Bessel beam at the vicinity of the TE Mie
resonances. (a) $ka=0.735$ around the bonding dipole resonance
$n=1$, (b) $ka=0.82$ around the anti-bonding dipole resonance
$n=1$, (c) $ka=1.0947$ around the bonding quadruple resonance
$n=2$, and (d) $ka=1.1783$ around the anti-bonding quadruple
resonance at $L=2a$. The solid line shows equilibrium positions of
spheres.}
\label{fig8}
\end{figure}

\section{Optical binding force  between two coaxial disks}
Distinct to the case of two spheres, two disks have two parameters
to vary, the aspect ratio and distance between disks. Even in an
isolated dielectric disk  the high-Q resonances can be achieved by
avoided crossing of the TE resonances of the same symmetry
relative to inversion of the disk's axis under variation of the
aspect ratio around $a/h=0.71$ as it was reported by Rybin {\it et
al} \cite{Rybin2017} and illustrated in Fig. \ref{fig10} (a) and
(b). While the resonances of the opposite symmetry in an isolated
disk plotted by solid and dash lines can not be coupled in single
disk. An example of this crossing is highlighted by circle in Fig.
\ref{fig10} (a). However, the presence of the second disk lifts
this symmetry restriction giving rise to a new series of avoided
crossings of resonances shown in Fig. \ref{fig10} (c)
\cite{Bulgakov2020}. In view of the OB force the most important is
the anti-bonding resonance which achieves unprecedent high $Q$
factor around 18000 as shown in Fig. \ref{fig10} (d). The reason
of such an extreme value is related to that the anti-bonding
resonant mode is close to the Mie resonant mode with extremely
large orbital index ($n=6$) of an effective sphere with the volume
equal $\pi(h+L)a^2$ shown in right bottom inset of Fig.
\ref{fig10} (c) \cite{Bulgakov2020}. That refers also to the
bonding resonant mode which is close to the Mie resonant mode with
$n=5$ shown in left bottom inset in Fig. \ref{fig10} (c).
Respectively we expect around the aspect ratio $a/h=1$ extremal
enhancement of OB, especially for the anti-bonding resonant mode
similar to Refs. \cite{Antonoyiannakis1997,Liu09,Zhang2014}. These
effective spheres are shown by white lines in bottom insets.
\begin{figure}[ht!]
\includegraphics[width=0.9\linewidth]{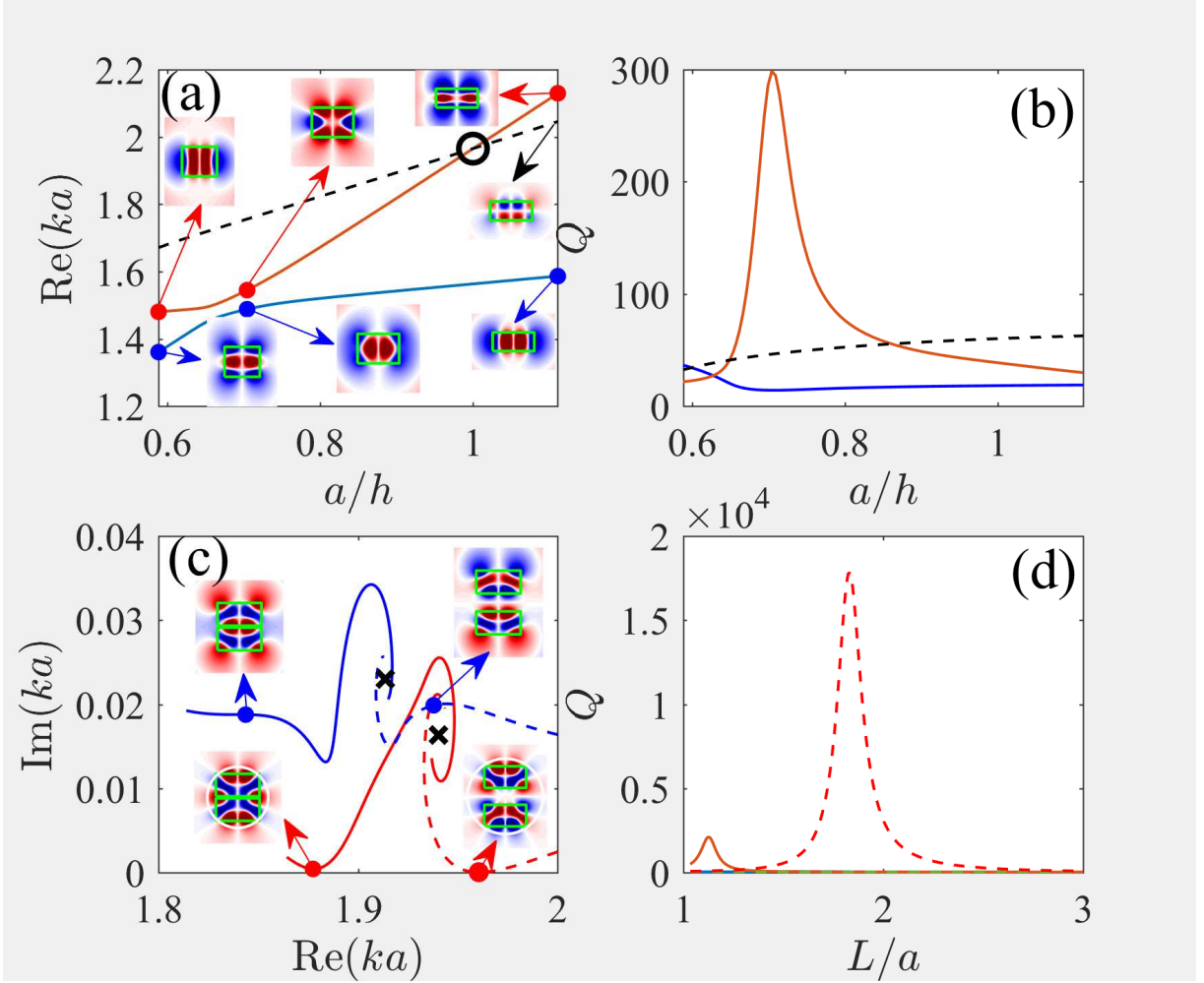}
\caption{(a) Avoided crossing of two TE resonances whose modes are
symmetric relative to $z\rightarrow -z$ for traversing over
distance and (b) their $Q$ factors versus the aspect ratio $a/h$
in isolated silicon disk. (c) Behavior of hybridized resonances
and (d) the $Q$ factor vs distance between centers of disks for
$a/h=1.003$. Insets show the profiles of tangential component of
electric field $E_{\phi}$.} \label{fig10}
\end{figure}


First, we consider a stability of single disk  at $r=0$. Numerical
calculations of forces by the centered Bessel beam and slightly
shifted beam relative to axis $r=0$ have shown that the position
of disk is stable at the symmetry axis at the vicinity of resonant
frequencies. That considerably simplifies the further calculation
of OB between two disks. The results of calculations of the OB are
presented in Figs. \ref{fig11} and \ref{fig13}.
\begin{figure}
\includegraphics*[width=8cm,clip=]{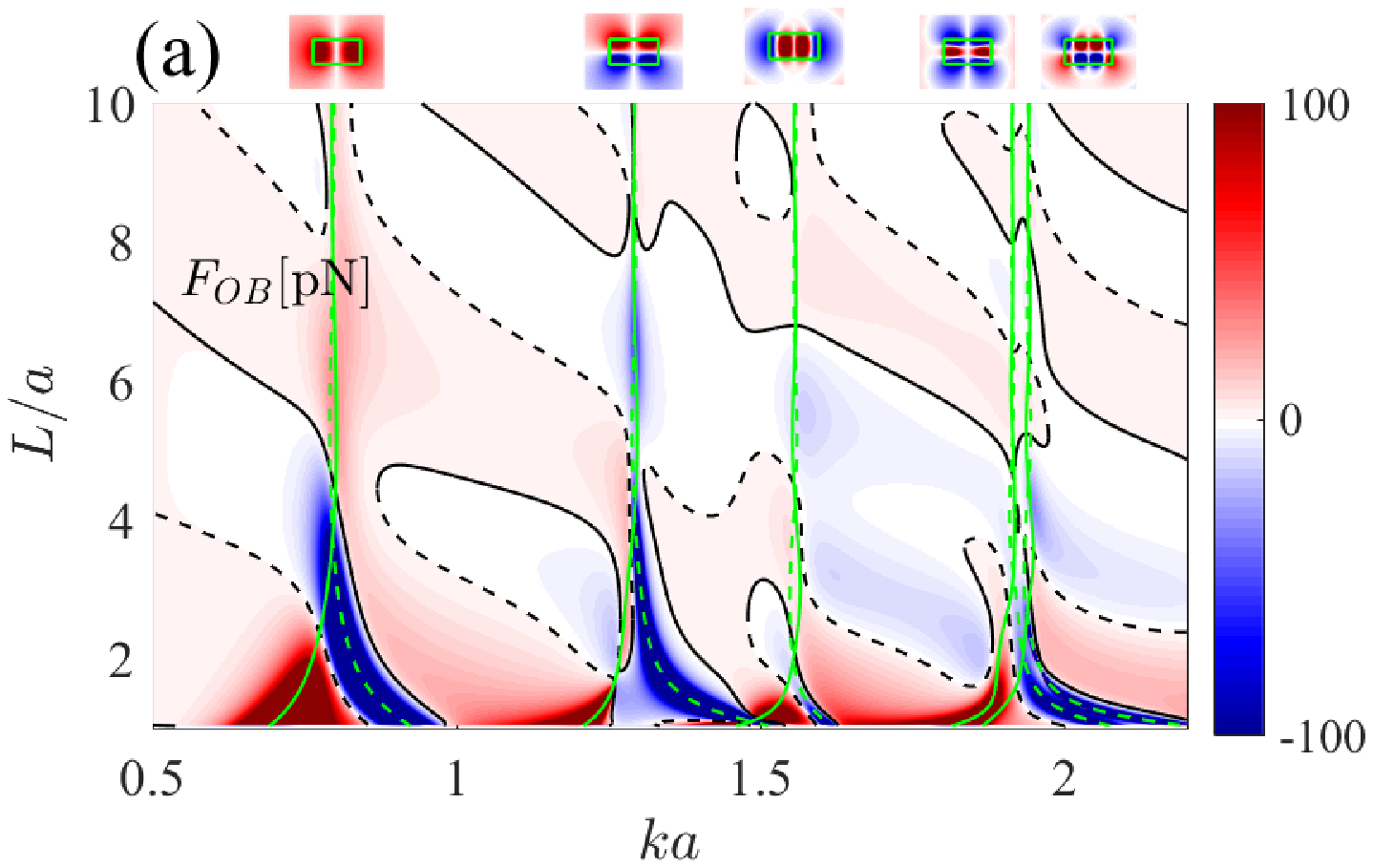}
\includegraphics*[width=8cm,clip=]{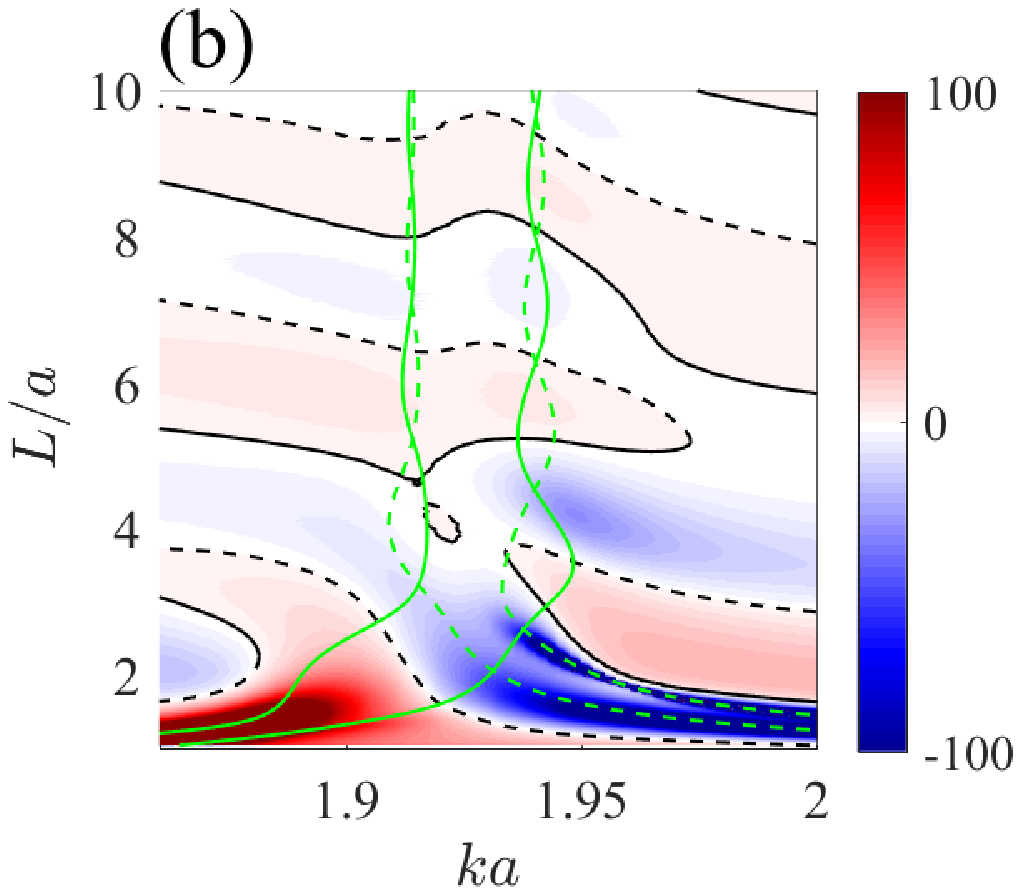}
\caption{The binding force between two disks vs the frequency and
distance between centers of disks for the Bessel beam with TE
polarization and $k_za=1/2$ where the disk  with $\epsilon=15$ has
the radius $a=0.5\mu m$. (b) zoomed versions. Black solid (dash)
lines show stable (unstable) configuration of disks. Light green
solid (dash) lines show bonding (symmetric) and anti-bonding (anti
symmetric) resonant frequencies of two disks vs the distance
between. } \label{fig11}
\end{figure}
\begin{figure}
\includegraphics*[width=8cm,clip=]{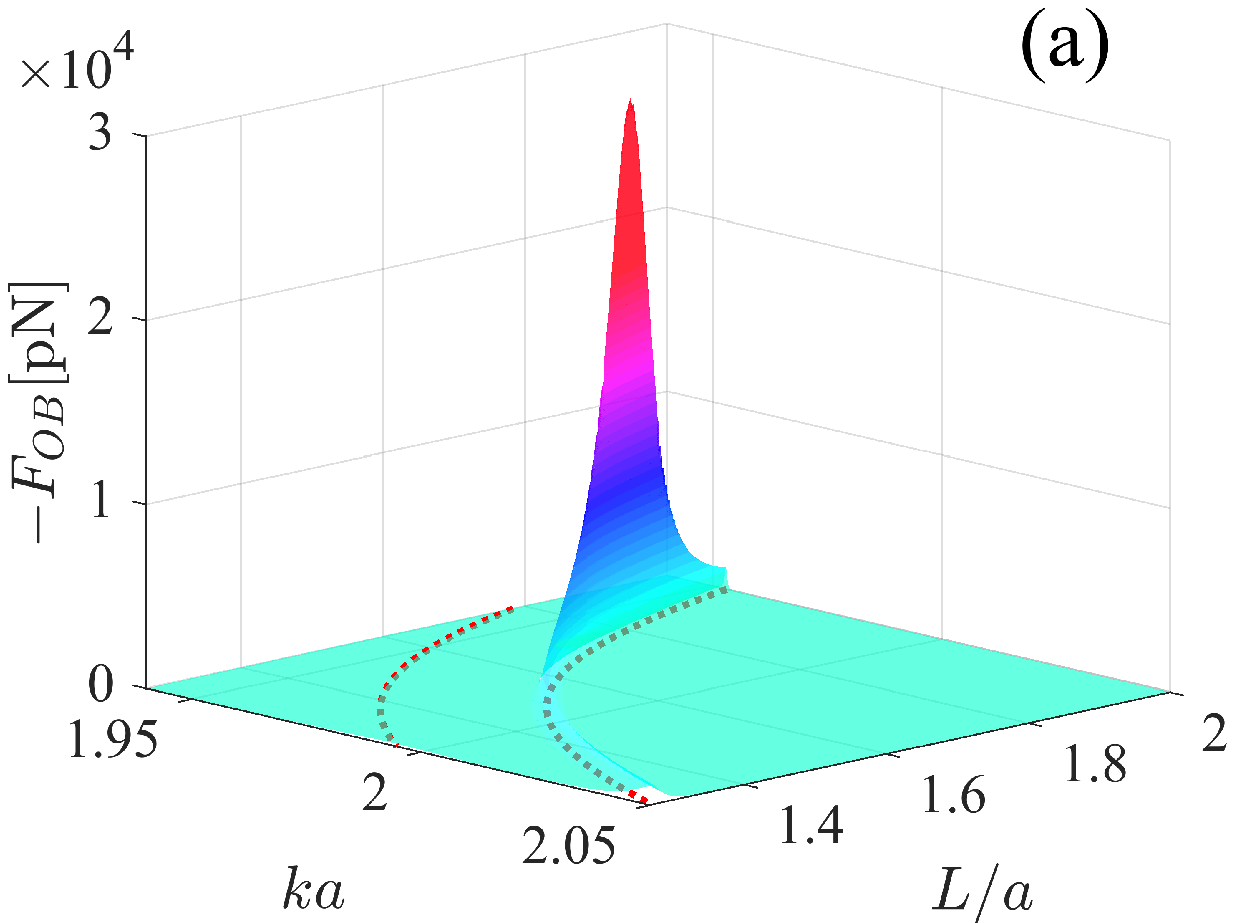}
\includegraphics*[width=8cm,clip=]{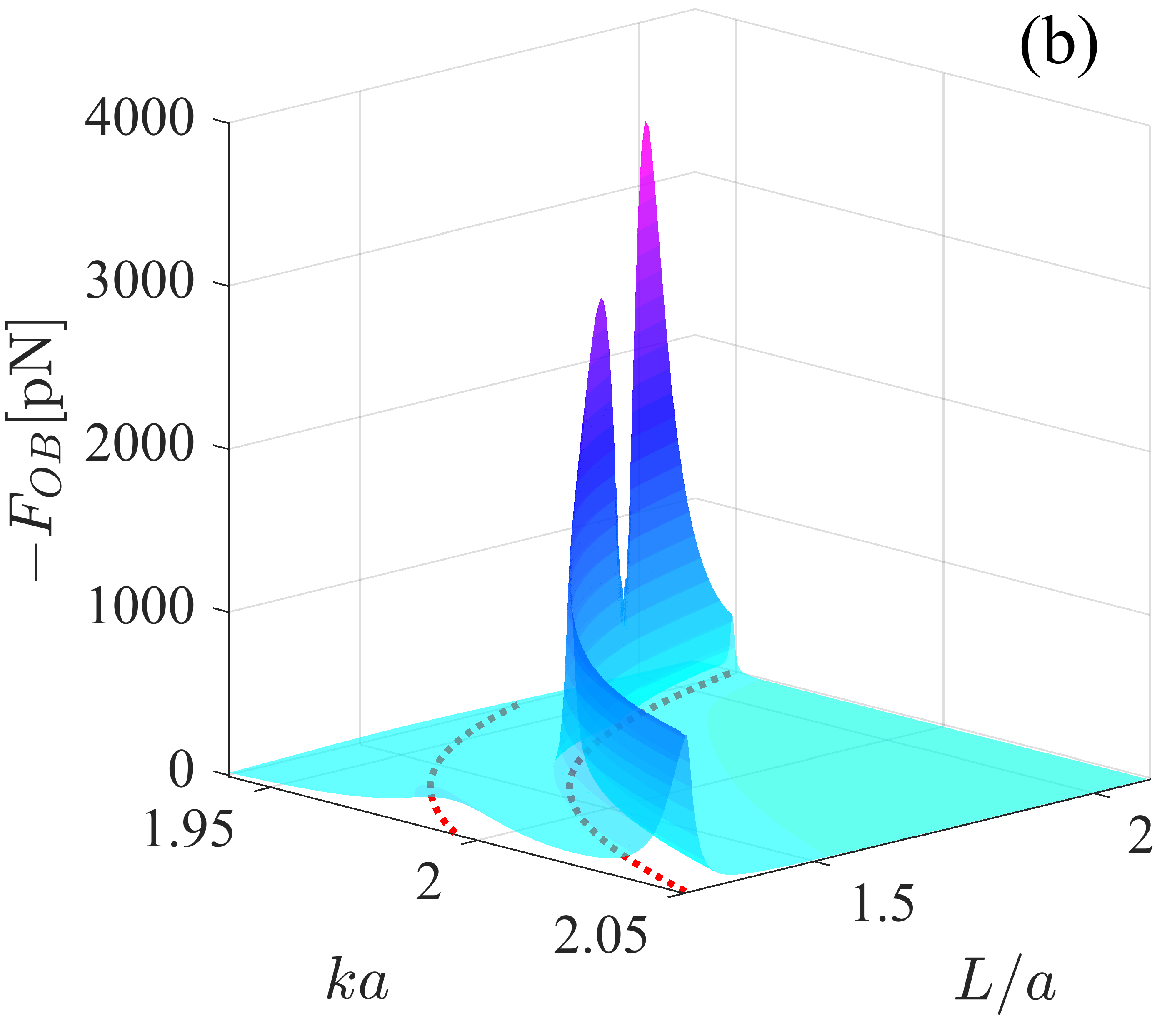}
\caption{The OB vs distance between centers of disks
at the vicinity of the anti-bonding  resonance marked in Fig.
\ref{fig10} (c) by closed circle $ka=1.95$ (a) $k_za=0.5$ and (b)
$k_za=1$. Solid line underneath shows resonant frequency vs
distance $L$ highlighted in Fig. \ref{fig11}.}
\label{fig13}
\end{figure}
Fig. \ref{fig13} demonstrates that indeed near the parameters of
extremely large peaks of the $Q$ factor we observe respectively
giant OB of order one  femto Newton. For the reader's convenience
we reproduce Fig. \ref{fig11} as surface in Fig. \ref{fig13} (a)
where one can see that giant OB is achieved around 30 femto
Newtons at $ka=1.97, L=1.85a, h=1.03a, k_za=0.5$. Fig. \ref{fig13}
(b) shows that this giant peak is split for $k_za=1$. It is
remarkable that the equilibrium distances between disks is
traversed close to the anti-bonding resonance shown by dotted
line. That situation was first reported for two dielectric slabs
which can move in waveguide that is equivalent to Fabry-Perot
resonator with high $Q$ resonances \cite{Sadreev16}.  Fig.
\ref{fig14} demonstrates that these giant peaks are easily
manipulated by small changes of parameters of the Bessel beam:
$k_za$ and frequency.
\begin{figure}
\includegraphics*[width=8cm,clip=]{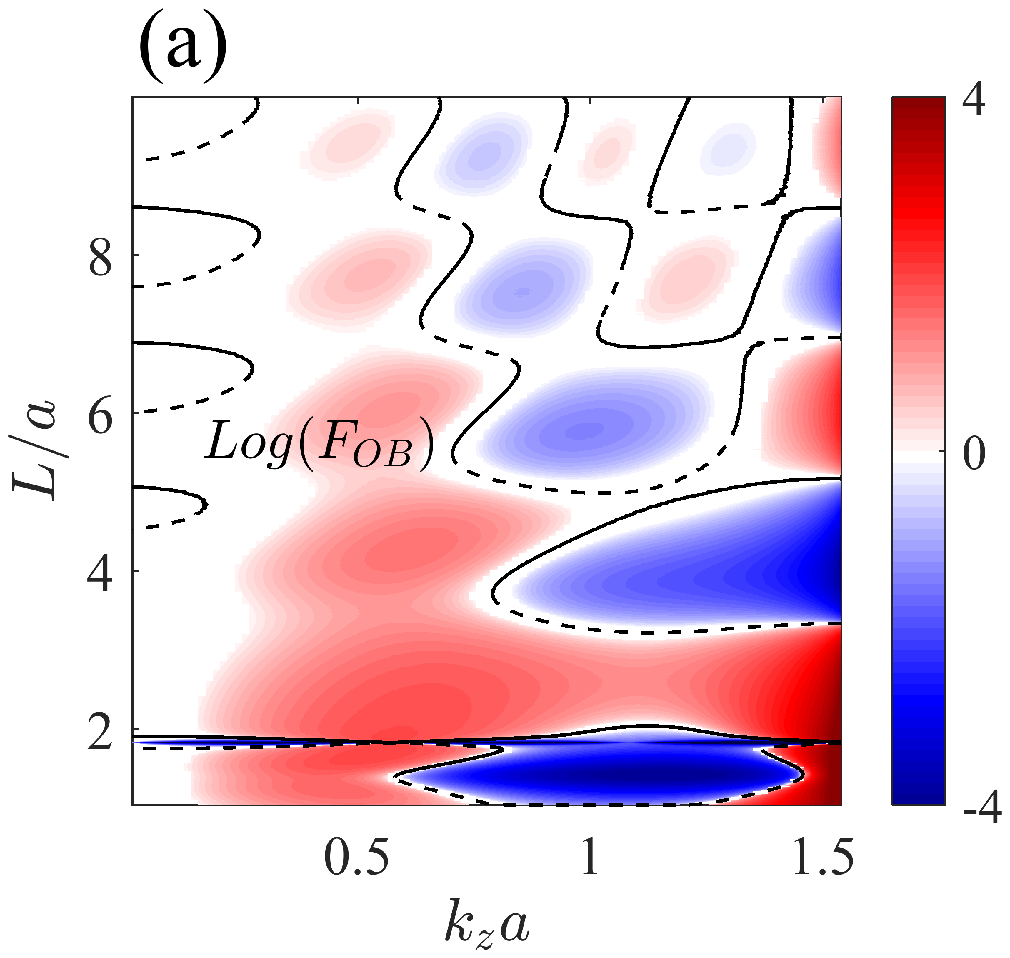}
\includegraphics*[width=8cm,clip=]{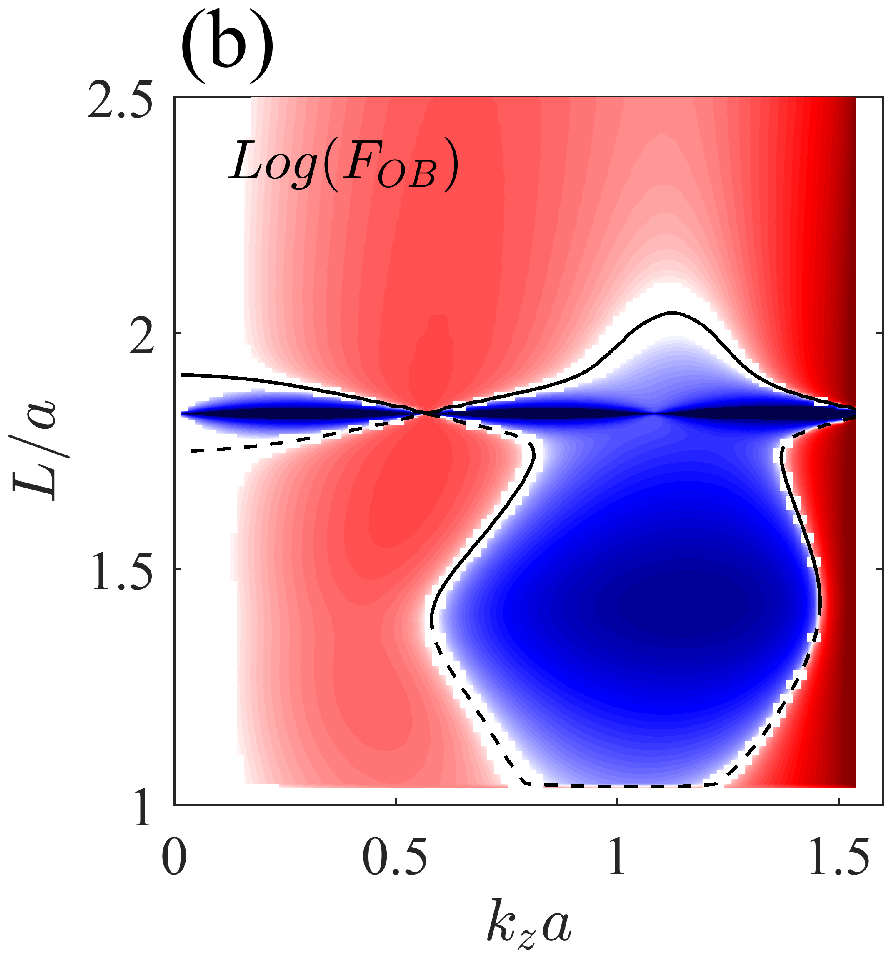}
\caption{(a) The OB vs distance between disks and
longitudinal wave vector of the Bessel beam $k_za$ at the vicinity
of the anti-bonding resonance marked in Fig. \ref{fig10} (c) by
closed circle $ka=1.95$. (b) Zoomed version of (a).}
\label{fig14}
\end{figure}

\section{Summary and conclusions}
In the present  paper we consider optical binding (OB) of
particles of micron size by illumination of dual counter
propagating Bessel beams. The case of two spheres owing to
formulas derived by Barton {\it et al} \cite{Barton1989} for
electromagnetic force acting on the isolated sphere  gives an
opportunity to derive analytical expressions for the OB force in
the resonant approximation. At large  distances the OB force
decays as inverse squared distance and has two periods of
oscillations (Eq. (\ref{OB})). For near distances the OB force can
be considerably enhanced up to order of one femto Newton. One of
the important and unexpected results of  the OB forces of spheres
is their decrease with growth of the order of the Mie resonances.
That is a result of competition of two types of couplings. The
first coupling of the Mie resonant modes of the sphere with the
radiation continua given by the vectorial spherical functions fast
falls with growth of the order of resonance giving rise to WGMs
with extremal $Q$ factors \cite{Lam1992,Gorodetsky1996}. However
the couplings of the Mie resonant modes with the incident Bessel
beams can decay even faster with the growth of the order of the
Mie resonant mode. Indeed, our calculations presented in Fig.
\ref{fig4} that the OB force is large only for the dipole and
quadruple Mie resonances.

  The case of coaxial disks brings a new aspect for the OB force related to
  the  extremely high $Q$ factor due to two-parametric
  avoided crossing of orthogonal resonances over aspect ratio and distance between the disks
  \cite{Bulgakov2020}. For the case of two coaxial silicon disks with micron diameter
illuminated by dual coaxial Bessel beams we demonstrate giant OB
force in few decades of femto Newtons in the  vicinity of
anti-bonding resonances. The corresponding anti-bonding resonant
mode of two disks is turned out to be close to the Mie resonant
mode with high orbital index $n=6$ of an effective sphere of the
volume $\pi a^2(h+L)$ \cite{Bulgakov2020}  with extremely high $Q$
factor. That allows to achieve giant OB force around a several
decades of femto Newtons.

There are three  important aspects of the OB force of two high
index dielectric particles. The first is a giant values of the
force around one nano Newtons for two spheres and a few decades of
nano Newtons for two coaxial disks illuminated by dual Bessel
beams with power $1mW/\mu m$. The second aspect is that the giant
OB forces are caused by resonant excitation of {\it subwavelenth}
resonant modes of particles. Potentially easy way for cardinal
manipulation of the OB force by a cross-section of the Bessel beam
constitutes the third aspect  of presented results.

\begin{acknowledgments}
The work was supported by Russian Foundation for Basic Research
projects No. 19-02-00055. AS thanks  Dmitrii Maksimov and Evgeny
Sherman for helpful discussions.
\end{acknowledgments}
\bibliography{sadreev}

\begin{thebibliography}{46}
\expandafter\ifx\csname natexlab\endcsname\relax\def\natexlab#1{#1}\fi
\expandafter\ifx\csname bibnamefont\endcsname\relax
  \def\bibnamefont#1{#1}\fi
\expandafter\ifx\csname bibfnamefont\endcsname\relax
  \def\bibfnamefont#1{#1}\fi
\expandafter\ifx\csname citenamefont\endcsname\relax
  \def\citenamefont#1{#1}\fi
\expandafter\ifx\csname url\endcsname\relax
  \def\url#1{\texttt{#1}}\fi
\expandafter\ifx\csname urlprefix\endcsname\relax\def\urlprefix{URL }\fi
\providecommand{\bibinfo}[2]{#2}
\providecommand{\eprint}[2][]{\url{#2}}

\bibitem[{\citenamefont{Ashkin et~al.}(1986)\citenamefont{Ashkin, Dziedzic,
  Bjorkholm, and Chu}}]{Ashkin1986}
\bibinfo{author}{\bibfnamefont{A.}~\bibnamefont{Ashkin}},
  \bibinfo{author}{\bibfnamefont{J.~M.} \bibnamefont{Dziedzic}},
  \bibinfo{author}{\bibfnamefont{J.~E.} \bibnamefont{Bjorkholm}},
  \bibnamefont{and} \bibinfo{author}{\bibfnamefont{S.}~\bibnamefont{Chu}},
  \bibinfo{journal}{Opt. Lett.} \textbf{\bibinfo{volume}{11}},
  \bibinfo{pages}{290} (\bibinfo{year}{1986}).

\bibitem[{\citenamefont{Burns et~al.}(1989)\citenamefont{Burns, Fournier, and
  Golovchenko}}]{Burns1989}
\bibinfo{author}{\bibfnamefont{M.~M.} \bibnamefont{Burns}},
  \bibinfo{author}{\bibfnamefont{J.-M.} \bibnamefont{Fournier}},
  \bibnamefont{and} \bibinfo{author}{\bibfnamefont{J.~A.}
  \bibnamefont{Golovchenko}}, \bibinfo{journal}{Phys. Rev. Lett.}
  \textbf{\bibinfo{volume}{63}}, \bibinfo{pages}{1233} (\bibinfo{year}{1989}).

\bibitem[{\citenamefont{Tatarkova et~al.}(2002)\citenamefont{Tatarkova,
  Carruthers, and Dholakia}}]{Tatarkova2002}
\bibinfo{author}{\bibfnamefont{S.~A.} \bibnamefont{Tatarkova}},
  \bibinfo{author}{\bibfnamefont{A.~E.} \bibnamefont{Carruthers}},
  \bibnamefont{and} \bibinfo{author}{\bibfnamefont{K.}~\bibnamefont{Dholakia}},
  \bibinfo{journal}{Phys. Rev. Lett.} \textbf{\bibinfo{volume}{89}}
  (\bibinfo{year}{2002}).

\bibitem[{\citenamefont{G{\'{o}}mez-Medina and
  S{\'{a}}enz}(2004)}]{Gomez-Medina2004}
\bibinfo{author}{\bibfnamefont{R.}~\bibnamefont{G{\'{o}}mez-Medina}}
  \bibnamefont{and} \bibinfo{author}{\bibfnamefont{J.~J.}
  \bibnamefont{S{\'{a}}enz}}, \bibinfo{journal}{Phys. Rev. Lett.}
  \textbf{\bibinfo{volume}{93}} (\bibinfo{year}{2004}).

\bibitem[{\citenamefont{Metzger
  et~al.}(2006{\natexlab{a}})\citenamefont{Metzger, Dholakia, and
  Wright}}]{Metzger2006}
\bibinfo{author}{\bibfnamefont{N.~K.} \bibnamefont{Metzger}},
  \bibinfo{author}{\bibfnamefont{K.}~\bibnamefont{Dholakia}}, \bibnamefont{and}
  \bibinfo{author}{\bibfnamefont{E.~M.} \bibnamefont{Wright}},
  \bibinfo{journal}{Phys. Rev. Lett.} \textbf{\bibinfo{volume}{96}}
  (\bibinfo{year}{2006}{\natexlab{a}}).

\bibitem[{\citenamefont{Metzger
  et~al.}(2006{\natexlab{b}})\citenamefont{Metzger, Wright, and
  Dholakia}}]{Metzger2006a}
\bibinfo{author}{\bibfnamefont{N.~K.} \bibnamefont{Metzger}},
  \bibinfo{author}{\bibfnamefont{E.~M.} \bibnamefont{Wright}},
  \bibnamefont{and} \bibinfo{author}{\bibfnamefont{K.}~\bibnamefont{Dholakia}},
  \bibinfo{journal}{New J. Phys.} \textbf{\bibinfo{volume}{8}},
  \bibinfo{pages}{139} (\bibinfo{year}{2006}{\natexlab{b}}).

\bibitem[{\citenamefont{Dholakia and Zem{\'{a}}nek}(2010)}]{Dholakia2010}
\bibinfo{author}{\bibfnamefont{K.}~\bibnamefont{Dholakia}} \bibnamefont{and}
  \bibinfo{author}{\bibfnamefont{P.}~\bibnamefont{Zem{\'{a}}nek}},
  \bibinfo{journal}{Rev. Mod. Phys.} \textbf{\bibinfo{volume}{82}},
  \bibinfo{pages}{1767} (\bibinfo{year}{2010}).

\bibitem[{\citenamefont{Bowman and Padgett}(2013)}]{Bowman2013}
\bibinfo{author}{\bibfnamefont{R.}~\bibnamefont{Bowman}} \bibnamefont{and}
  \bibinfo{author}{\bibfnamefont{M.}~\bibnamefont{Padgett}},
  \bibinfo{journal}{Rep. Progr. Phys.} \textbf{\bibinfo{volume}{76}},
  \bibinfo{pages}{026401} (\bibinfo{year}{2013}).

\bibitem[{\citenamefont{Thanopulos et~al.}(2017)\citenamefont{Thanopulos,
  Luckhaus, and Signorell}}]{Thanopulos2017}
\bibinfo{author}{\bibfnamefont{I.}~\bibnamefont{Thanopulos}},
  \bibinfo{author}{\bibfnamefont{D.}~\bibnamefont{Luckhaus}}, \bibnamefont{and}
  \bibinfo{author}{\bibfnamefont{R.}~\bibnamefont{Signorell}},
  \bibinfo{journal}{Phys. Rev. A} \textbf{\bibinfo{volume}{95}}
  (\bibinfo{year}{2017}).

\bibitem[{\citenamefont{Chaumet and Nieto-Vesperinas}(2001)}]{Chaumet2001}
\bibinfo{author}{\bibfnamefont{P.~C.} \bibnamefont{Chaumet}} \bibnamefont{and}
  \bibinfo{author}{\bibfnamefont{M.}~\bibnamefont{Nieto-Vesperinas}},
  \bibinfo{journal}{Phys. Rev. B} \textbf{\bibinfo{volume}{64}}
  (\bibinfo{year}{2001}).

\bibitem[{\citenamefont{Ng et~al.}(2005)\citenamefont{Ng, Lin, Chan, and
  Sheng}}]{Ng2005}
\bibinfo{author}{\bibfnamefont{J.}~\bibnamefont{Ng}},
  \bibinfo{author}{\bibfnamefont{Z.}~\bibnamefont{Lin}},
  \bibinfo{author}{\bibfnamefont{C.}~\bibnamefont{Chan}}, \bibnamefont{and}
  \bibinfo{author}{\bibfnamefont{P.}~\bibnamefont{Sheng}},
  \bibinfo{journal}{Phys. Rev. B} \textbf{\bibinfo{volume}{72}}
  (\bibinfo{year}{2005}).

\bibitem[{\citenamefont{{\v{C}}i{\v{z}}m{\'{a}}r
  et~al.}(2006)\citenamefont{{\v{C}}i{\v{z}}m{\'{a}}r, Koll{\'{a}}rov{\'{a}},
  Bouchal, and Zem{\'{a}}nek}}]{Cizmar2006}
\bibinfo{author}{\bibfnamefont{T.}~\bibnamefont{{\v{C}}i{\v{z}}m{\'{a}}r}},
  \bibinfo{author}{\bibfnamefont{V.}~\bibnamefont{Koll{\'{a}}rov{\'{a}}}},
  \bibinfo{author}{\bibfnamefont{Z.}~\bibnamefont{Bouchal}}, \bibnamefont{and}
  \bibinfo{author}{\bibfnamefont{P.}~\bibnamefont{Zem{\'{a}}nek}},
  \bibinfo{journal}{New J. Phys.} \textbf{\bibinfo{volume}{8}},
  \bibinfo{pages}{43} (\bibinfo{year}{2006}).

\bibitem[{\citenamefont{Thanopulos et~al.}(2014)\citenamefont{Thanopulos,
  Luckhaus, Preston, and Signorell}}]{Thanopulos2014}
\bibinfo{author}{\bibfnamefont{I.}~\bibnamefont{Thanopulos}},
  \bibinfo{author}{\bibfnamefont{D.}~\bibnamefont{Luckhaus}},
  \bibinfo{author}{\bibfnamefont{T.}~\bibnamefont{Preston}}, \bibnamefont{and}
  \bibinfo{author}{\bibfnamefont{R.}~\bibnamefont{Signorell}},
  \bibinfo{journal}{J. Appl. Phys.} \textbf{\bibinfo{volume}{115}},
  \bibinfo{pages}{154304} (\bibinfo{year}{2014}).

\bibitem[{\citenamefont{Antonoyiannakis and
  Pendry}(1997)}]{Antonoyiannakis1997}
\bibinfo{author}{\bibfnamefont{M.~I.} \bibnamefont{Antonoyiannakis}}
  \bibnamefont{and} \bibinfo{author}{\bibfnamefont{J.~B.}
  \bibnamefont{Pendry}}, \bibinfo{journal}{Europhys. Lett.}
  \textbf{\bibinfo{volume}{40}}, \bibinfo{pages}{613} (\bibinfo{year}{1997}).

\bibitem[{\citenamefont{Liu et~al.}(2009)\citenamefont{Liu, Povinelli, and
  Fan}}]{Liu09}
\bibinfo{author}{\bibfnamefont{V.}~\bibnamefont{Liu}},
  \bibinfo{author}{\bibfnamefont{M.}~\bibnamefont{Povinelli}},
  \bibnamefont{and} \bibinfo{author}{\bibfnamefont{S.}~\bibnamefont{Fan}},
  \bibinfo{journal}{Opt. Express} \textbf{\bibinfo{volume}{17}}
  (\bibinfo{year}{2009}).

\bibitem[{\citenamefont{Zhang et~al.}(2014)\citenamefont{Zhang, MacDonald, and
  Zheludev}}]{Zhang2014}
\bibinfo{author}{\bibfnamefont{J.}~\bibnamefont{Zhang}},
  \bibinfo{author}{\bibfnamefont{K.}~\bibnamefont{MacDonald}},
  \bibnamefont{and} \bibinfo{author}{\bibfnamefont{N.}~\bibnamefont{Zheludev}},
  \bibinfo{journal}{Opt. Lett.} \textbf{\bibinfo{volume}{39}},
  \bibinfo{pages}{4883} (\bibinfo{year}{2014}).

\bibitem[{\citenamefont{Taghizadeh and Chung}(2017)}]{Taghizadeh2017}
\bibinfo{author}{\bibfnamefont{A.}~\bibnamefont{Taghizadeh}} \bibnamefont{and}
  \bibinfo{author}{\bibfnamefont{I.-S.} \bibnamefont{Chung}},
  \bibinfo{journal}{Appl. Phys. Lett.} \textbf{\bibinfo{volume}{111}},
  \bibinfo{pages}{031114} (\bibinfo{year}{2017}).

\bibitem[{\citenamefont{Sadrieva et~al.}(2019)\citenamefont{Sadrieva, Belyakov,
  Balezin, Kapitanova, Nenasheva, Sadreev, and Bogdanov}}]{Sadrieva2019}
\bibinfo{author}{\bibfnamefont{Z.~F.} \bibnamefont{Sadrieva}},
  \bibinfo{author}{\bibfnamefont{M.~A.} \bibnamefont{Belyakov}},
  \bibinfo{author}{\bibfnamefont{M.~A.} \bibnamefont{Balezin}},
  \bibinfo{author}{\bibfnamefont{P.~V.} \bibnamefont{Kapitanova}},
  \bibinfo{author}{\bibfnamefont{E.~A.} \bibnamefont{Nenasheva}},
  \bibinfo{author}{\bibfnamefont{A.~F.} \bibnamefont{Sadreev}},
  \bibnamefont{and} \bibinfo{author}{\bibfnamefont{A.~A.}
  \bibnamefont{Bogdanov}}, \bibinfo{journal}{Phys. Rev. A}
  \textbf{\bibinfo{volume}{99}}, \bibinfo{pages}{053804}
  (\bibinfo{year}{2019}).

\bibitem[{\citenamefont{Bulgakov and Sadreev}(2019)}]{Bulgakov2019a}
\bibinfo{author}{\bibfnamefont{E.}~\bibnamefont{Bulgakov}} \bibnamefont{and}
  \bibinfo{author}{\bibfnamefont{A.}~\bibnamefont{Sadreev}},
  \bibinfo{journal}{Phys. Rev. A} \textbf{\bibinfo{volume}{99}},
  \bibinfo{pages}{033851} (\bibinfo{year}{2019}).

\bibitem[{\citenamefont{Povinelli et~al.}(2005)\citenamefont{Povinelli,
  Johnson, Loncar, Ibanescu, Smythe, Capasso, and
  Joannopoulos}}]{Povinelli2005}
\bibinfo{author}{\bibfnamefont{M.~L.} \bibnamefont{Povinelli}},
  \bibinfo{author}{\bibfnamefont{S.}~\bibnamefont{Johnson}},
  \bibinfo{author}{\bibfnamefont{M.}~\bibnamefont{Loncar}},
  \bibinfo{author}{\bibfnamefont{M.}~\bibnamefont{Ibanescu}},
  \bibinfo{author}{\bibfnamefont{E.}~\bibnamefont{Smythe}},
  \bibinfo{author}{\bibfnamefont{F.}~\bibnamefont{Capasso}}, \bibnamefont{and}
  \bibinfo{author}{\bibfnamefont{J.~D.} \bibnamefont{Joannopoulos}},
  \bibinfo{journal}{Optics Express} \textbf{\bibinfo{volume}{13}},
  \bibinfo{pages}{8286} (\bibinfo{year}{2005}).

\bibitem[{\citenamefont{Benyoucef et~al.}(2011)\citenamefont{Benyoucef, Shim,
  Wiersig, and Schmidt}}]{Benyoucef2011}
\bibinfo{author}{\bibfnamefont{M.}~\bibnamefont{Benyoucef}},
  \bibinfo{author}{\bibfnamefont{J.-B.} \bibnamefont{Shim}},
  \bibinfo{author}{\bibfnamefont{J.}~\bibnamefont{Wiersig}}, \bibnamefont{and}
  \bibinfo{author}{\bibfnamefont{O.~G.} \bibnamefont{Schmidt}},
  \bibinfo{journal}{Opt. Lett.} \textbf{\bibinfo{volume}{36}},
  \bibinfo{pages}{1317} (\bibinfo{year}{2011}).

\bibitem[{\citenamefont{Bulgakov et~al.}(2020)\citenamefont{Bulgakov, Pichugin,
  and Sadreev}}]{Bulgakov2020}
\bibinfo{author}{\bibfnamefont{E.~N.} \bibnamefont{Bulgakov}},
  \bibinfo{author}{\bibfnamefont{K.~N.} \bibnamefont{Pichugin}},
  \bibnamefont{and} \bibinfo{author}{\bibfnamefont{A.~F.}
  \bibnamefont{Sadreev}} (\bibinfo{year}{2020}), \eprint{2005.05554v1}.

\bibitem[{\citenamefont{Barton et~al.}(1989)\citenamefont{Barton, Alexander,
  and Schaub}}]{Barton1989}
\bibinfo{author}{\bibfnamefont{J.~P.} \bibnamefont{Barton}},
  \bibinfo{author}{\bibfnamefont{D.~R.} \bibnamefont{Alexander}},
  \bibnamefont{and} \bibinfo{author}{\bibfnamefont{S.~A.}
  \bibnamefont{Schaub}}, \bibinfo{journal}{J. Appl. Phys.}
  \textbf{\bibinfo{volume}{66}}, \bibinfo{pages}{4594} (\bibinfo{year}{1989}).

\bibitem[{\citenamefont{Kar{\'{a}}sek and Zem{\'{a}}nek}(2007)}]{Karasek2007}
\bibinfo{author}{\bibfnamefont{V.}~\bibnamefont{Kar{\'{a}}sek}}
  \bibnamefont{and}
  \bibinfo{author}{\bibfnamefont{P.}~\bibnamefont{Zem{\'{a}}nek}},
  \bibinfo{journal}{Journal of Optics A: Pure and Applied Optics}
  \textbf{\bibinfo{volume}{9}}, \bibinfo{pages}{S215} (\bibinfo{year}{2007}).

\bibitem[{\citenamefont{Kar{\'{a}}sek et~al.}(2009)\citenamefont{Kar{\'{a}}sek,
  Brzobohat{\'{y}}, and Zem{\'{a}}nek}}]{Karasek2009}
\bibinfo{author}{\bibfnamefont{V.}~\bibnamefont{Kar{\'{a}}sek}},
  \bibinfo{author}{\bibfnamefont{O.}~\bibnamefont{Brzobohat{\'{y}}}},
  \bibnamefont{and}
  \bibinfo{author}{\bibfnamefont{P.}~\bibnamefont{Zem{\'{a}}nek}},
  \bibinfo{journal}{J. Optics A: Pure and Appl. Optics}
  \textbf{\bibinfo{volume}{11}}, \bibinfo{pages}{034009}
  (\bibinfo{year}{2009}).

\bibitem[{\citenamefont{Zhu et~al.}(2015)\citenamefont{Zhu, Wu, Li, and
  Shang}}]{Zhu2015}
\bibinfo{author}{\bibfnamefont{Y.}~\bibnamefont{Zhu}},
  \bibinfo{author}{\bibfnamefont{Z.}~\bibnamefont{Wu}},
  \bibinfo{author}{\bibfnamefont{Z.}~\bibnamefont{Li}}, \bibnamefont{and}
  \bibinfo{author}{\bibfnamefont{Q.}~\bibnamefont{Shang}},
  \bibinfo{journal}{Procedia Engineering} \textbf{\bibinfo{volume}{102}},
  \bibinfo{pages}{329} (\bibinfo{year}{2015}).

\bibitem[{\citenamefont{Deng et~al.}(2018)\citenamefont{Deng, Liu, Panmai, and
  Lan}}]{Deng2018}
\bibinfo{author}{\bibfnamefont{F.}~\bibnamefont{Deng}},
  \bibinfo{author}{\bibfnamefont{H.}~\bibnamefont{Liu}},
  \bibinfo{author}{\bibfnamefont{M.}~\bibnamefont{Panmai}}, \bibnamefont{and}
  \bibinfo{author}{\bibfnamefont{S.}~\bibnamefont{Lan}}, \bibinfo{journal}{Opt.
  Express} \textbf{\bibinfo{volume}{26}}, \bibinfo{pages}{20051}
  (\bibinfo{year}{2018}).

\bibitem[{\citenamefont{Milne et~al.}(2007)\citenamefont{Milne, Dholakia,
  McGloin, Volke-Sepulveda, and Zem{\'{a}}nek}}]{Milne2007}
\bibinfo{author}{\bibfnamefont{G.}~\bibnamefont{Milne}},
  \bibinfo{author}{\bibfnamefont{K.}~\bibnamefont{Dholakia}},
  \bibinfo{author}{\bibfnamefont{D.}~\bibnamefont{McGloin}},
  \bibinfo{author}{\bibfnamefont{K.}~\bibnamefont{Volke-Sepulveda}},
  \bibnamefont{and}
  \bibinfo{author}{\bibfnamefont{P.}~\bibnamefont{Zem{\'{a}}nek}},
  \bibinfo{journal}{Opt. Express} \textbf{\bibinfo{volume}{15}},
  \bibinfo{pages}{13972} (\bibinfo{year}{2007}).

\bibitem[{\citenamefont{L.D.Landau and E.M.Lifshitz}(1960)}]{LL}
\bibinfo{author}{\bibnamefont{L.D.Landau}} \bibnamefont{and}
  \bibinfo{author}{\bibnamefont{E.M.Lifshitz}},
  \emph{\bibinfo{title}{Electrodynamics of Continuous Media}}
  (\bibinfo{publisher}{Pergamon, New York}, \bibinfo{year}{1960}).

\bibitem[{\citenamefont{Antonoyiannakis and Pendry}(1999)}]{Antonoyiannakis99}
\bibinfo{author}{\bibfnamefont{M.~I.} \bibnamefont{Antonoyiannakis}}
  \bibnamefont{and} \bibinfo{author}{\bibfnamefont{J.~B.}
  \bibnamefont{Pendry}}, \bibinfo{journal}{Phys. Rev. B}
  \textbf{\bibinfo{volume}{60}}, \bibinfo{pages}{2363} (\bibinfo{year}{1999}).

\bibitem[{\citenamefont{Chen et~al.}(2009)\citenamefont{Chen, Ng, Liu, and
  Lin}}]{Chen2009}
\bibinfo{author}{\bibfnamefont{J.}~\bibnamefont{Chen}},
  \bibinfo{author}{\bibfnamefont{J.}~\bibnamefont{Ng}},
  \bibinfo{author}{\bibfnamefont{S.}~\bibnamefont{Liu}}, \bibnamefont{and}
  \bibinfo{author}{\bibfnamefont{Z.}~\bibnamefont{Lin}},
  \bibinfo{journal}{Phys. Rev. E} \textbf{\bibinfo{volume}{80}}
  (\bibinfo{year}{2009}).

\bibitem[{\citenamefont{Wang et~al.}(2013)\citenamefont{Wang, Chen, Liu, and
  Lin}}]{Wang2013}
\bibinfo{author}{\bibfnamefont{N.}~\bibnamefont{Wang}},
  \bibinfo{author}{\bibfnamefont{J.}~\bibnamefont{Chen}},
  \bibinfo{author}{\bibfnamefont{S.}~\bibnamefont{Liu}}, \bibnamefont{and}
  \bibinfo{author}{\bibfnamefont{Z.}~\bibnamefont{Lin}},
  \bibinfo{journal}{Phys. Rev. A} \textbf{\bibinfo{volume}{87}}
  (\bibinfo{year}{2013}).

\bibitem[{\citenamefont{Song et~al.}(2014)\citenamefont{Song, Wang, Lu, and
  Lin}}]{Song2014}
\bibinfo{author}{\bibfnamefont{S.}~\bibnamefont{Song}},
  \bibinfo{author}{\bibfnamefont{N.}~\bibnamefont{Wang}},
  \bibinfo{author}{\bibfnamefont{W.}~\bibnamefont{Lu}}, \bibnamefont{and}
  \bibinfo{author}{\bibfnamefont{Z.}~\bibnamefont{Lin}}, \bibinfo{journal}{J.
  Opt. Soc. Am. A} \textbf{\bibinfo{volume}{31}}, \bibinfo{pages}{2192}
  (\bibinfo{year}{2014}).

\bibitem[{\citenamefont{Kiselev and Plutenko}(2016)}]{Kiselev2016}
\bibinfo{author}{\bibfnamefont{A.~D.} \bibnamefont{Kiselev}} \bibnamefont{and}
  \bibinfo{author}{\bibfnamefont{D.~O.} \bibnamefont{Plutenko}},
  \bibinfo{journal}{Phys. Rev. A} \textbf{\bibinfo{volume}{94}}
  (\bibinfo{year}{2016}).

\bibitem[{\citenamefont{Neves and Cesar}(2019)}]{Neves2019}
\bibinfo{author}{\bibfnamefont{A.}~\bibnamefont{Neves}} \bibnamefont{and}
  \bibinfo{author}{\bibfnamefont{C.}~\bibnamefont{Cesar}}, \bibinfo{journal}{J.
  Opt. Soc. Am. B} \textbf{\bibinfo{volume}{36}}, \bibinfo{pages}{1525}
  (\bibinfo{year}{2019}).

\bibitem[{\citenamefont{Stratton}(1941)}]{Stratton}
\bibinfo{author}{\bibfnamefont{J.}~\bibnamefont{Stratton}},
  \emph{\bibinfo{title}{Electromagnetic theory}}
  (\bibinfo{publisher}{McGraw-Hill Book Company, Inc.}, \bibinfo{year}{1941}).

\bibitem[{\citenamefont{Linton et~al.}(2013)\citenamefont{Linton, Zalipaev, and
  Thompson}}]{Linton2013}
\bibinfo{author}{\bibfnamefont{C.}~\bibnamefont{Linton}},
  \bibinfo{author}{\bibfnamefont{V.}~\bibnamefont{Zalipaev}}, \bibnamefont{and}
  \bibinfo{author}{\bibfnamefont{I.}~\bibnamefont{Thompson}},
  \bibinfo{journal}{Wave Motion} \textbf{\bibinfo{volume}{50}},
  \bibinfo{pages}{29} (\bibinfo{year}{2013}).

\bibitem[{\citenamefont{Mackowski}(1991)}]{Mackowski1991}
\bibinfo{author}{\bibfnamefont{D.}~\bibnamefont{Mackowski}},
  \bibinfo{journal}{Proceedings of the Royal Society of London. Series A: Math.
  and Phys. Sciences} \textbf{\bibinfo{volume}{433}}, \bibinfo{pages}{599}
  (\bibinfo{year}{1991}).

\bibitem[{\citenamefont{Pichugin and Sadreev}(2019)}]{Pichugin2019}
\bibinfo{author}{\bibfnamefont{K.~N.} \bibnamefont{Pichugin}} \bibnamefont{and}
  \bibinfo{author}{\bibfnamefont{A.~F.} \bibnamefont{Sadreev}},
  \bibinfo{journal}{J. Appl. Phys.} \textbf{\bibinfo{volume}{126}},
  \bibinfo{pages}{093105} (\bibinfo{year}{2019}).

\bibitem[{\citenamefont{Song et~al.}(2019)\citenamefont{Song, Zhao, Liu, Chai,
  He, Xiang, Han, and Zi}}]{Song2019}
\bibinfo{author}{\bibfnamefont{Q.}~\bibnamefont{Song}},
  \bibinfo{author}{\bibfnamefont{M.}~\bibnamefont{Zhao}},
  \bibinfo{author}{\bibfnamefont{L.}~\bibnamefont{Liu}},
  \bibinfo{author}{\bibfnamefont{J.}~\bibnamefont{Chai}},
  \bibinfo{author}{\bibfnamefont{G.}~\bibnamefont{He}},
  \bibinfo{author}{\bibfnamefont{H.}~\bibnamefont{Xiang}},
  \bibinfo{author}{\bibfnamefont{D.}~\bibnamefont{Han}}, \bibnamefont{and}
  \bibinfo{author}{\bibfnamefont{J.}~\bibnamefont{Zi}}, \bibinfo{journal}{Phys.
  Rev. A} \textbf{\bibinfo{volume}{100}} (\bibinfo{year}{2019}).

\bibitem[{\citenamefont{van~de Nes and Torok}(2007)}]{Nes2007}
\bibinfo{author}{\bibfnamefont{A.~S.} \bibnamefont{van~de Nes}}
  \bibnamefont{and} \bibinfo{author}{\bibfnamefont{P.}~\bibnamefont{Torok}},
  \bibinfo{journal}{Opt. Express} \textbf{\bibinfo{volume}{15}},
  \bibinfo{pages}{13360} (\bibinfo{year}{2007}).

\bibitem[{\citenamefont{Jiang et~al.}(2012)\citenamefont{Jiang, Shao, Qu, Ou,
  and Hua}}]{Jiang2012}
\bibinfo{author}{\bibfnamefont{Y.}~\bibnamefont{Jiang}},
  \bibinfo{author}{\bibfnamefont{Y.}~\bibnamefont{Shao}},
  \bibinfo{author}{\bibfnamefont{X.}~\bibnamefont{Qu}},
  \bibinfo{author}{\bibfnamefont{J.}~\bibnamefont{Ou}}, \bibnamefont{and}
  \bibinfo{author}{\bibfnamefont{H.}~\bibnamefont{Hua}}, \bibinfo{journal}{J.
  Opt.} \textbf{\bibinfo{volume}{14}}, \bibinfo{pages}{125709}
  (\bibinfo{year}{2012}).

\bibitem[{\citenamefont{Lam et~al.}(1992)\citenamefont{Lam, Leung, and
  Young}}]{Lam1992}
\bibinfo{author}{\bibfnamefont{C.~C.} \bibnamefont{Lam}},
  \bibinfo{author}{\bibfnamefont{P.~T.} \bibnamefont{Leung}}, \bibnamefont{and}
  \bibinfo{author}{\bibfnamefont{K.}~\bibnamefont{Young}}, \bibinfo{journal}{J.
  Opt. Soc. Am. B} \textbf{\bibinfo{volume}{9}}, \bibinfo{pages}{1585}
  (\bibinfo{year}{1992}).

\bibitem[{\citenamefont{Gorodetsky et~al.}(1996)\citenamefont{Gorodetsky,
  Savchenkov, and Ilchenko}}]{Gorodetsky1996}
\bibinfo{author}{\bibfnamefont{M.~L.} \bibnamefont{Gorodetsky}},
  \bibinfo{author}{\bibfnamefont{A.~A.} \bibnamefont{Savchenkov}},
  \bibnamefont{and} \bibinfo{author}{\bibfnamefont{V.~S.}
  \bibnamefont{Ilchenko}}, \bibinfo{journal}{Opt. Lett.}
  \textbf{\bibinfo{volume}{21}}, \bibinfo{pages}{453} (\bibinfo{year}{1996}).

\bibitem[{\citenamefont{Rybin et~al.}(2017)\citenamefont{Rybin, Koshelev,
  Sadrieva, Samusev, Bogdanov, Limonov, and Kivshar}}]{Rybin2017}
\bibinfo{author}{\bibfnamefont{M.}~\bibnamefont{Rybin}},
  \bibinfo{author}{\bibfnamefont{K.}~\bibnamefont{Koshelev}},
  \bibinfo{author}{\bibfnamefont{Z.}~\bibnamefont{Sadrieva}},
  \bibinfo{author}{\bibfnamefont{K.}~\bibnamefont{Samusev}},
  \bibinfo{author}{\bibfnamefont{A.}~\bibnamefont{Bogdanov}},
  \bibinfo{author}{\bibfnamefont{M.}~\bibnamefont{Limonov}}, \bibnamefont{and}
  \bibinfo{author}{\bibfnamefont{Y.}~\bibnamefont{Kivshar}},
  \bibinfo{journal}{Phys. Rev. Lett.} \textbf{\bibinfo{volume}{119}},
  \bibinfo{pages}{243901} (\bibinfo{year}{2017}).

\bibitem[{\citenamefont{Sadreev and Sherman}(2016)}]{Sadreev16}
\bibinfo{author}{\bibfnamefont{A.}~\bibnamefont{Sadreev}} \bibnamefont{and}
  \bibinfo{author}{\bibfnamefont{E.~Y.} \bibnamefont{Sherman}},
  \bibinfo{journal}{Phys. Rev. A} \textbf{\bibinfo{volume}{94}},
  \bibinfo{pages}{033820} (\bibinfo{year}{2016}).

\end{thebibliography}
\end{document}